\documentclass[aps,prx,showkeys,superscriptaddress,reprint]{revtex4-2}

\usepackage[utf8]{inputenc}
\usepackage[T1]{fontenc}
\usepackage{amsmath}
\usepackage{amssymb}
\usepackage{amsfonts}
\usepackage{upgreek}
\usepackage{mathtools}
\usepackage{siunitx}
\usepackage{float}
\usepackage{graphicx}
\usepackage{tabularx}
\usepackage[labelfont=bf]{caption}
\usepackage{booktabs} % for better rules for tables
\usepackage{threeparttable} % for notes under tables

\usepackage[hidelinks]{hyperref}
\usepackage[normalem]{ulem}
\usepackage{subcaption}

\begin{document}

\title{Machine Learning Enhanced Calculation of Quantum-Classical Binding Free Energies}

\author{Moritz Bensberg}
\email{Authors contributed equally.}
\affiliation{ETH Zurich, Department of Chemistry and Applied Biosciences, Vladimir-Prelog-Weg 2, 8093 Zurich, Switzerland.}
\author{Marco Eckhoff}
\email{Authors contributed equally.}
\affiliation{ETH Zurich, Department of Chemistry and Applied Biosciences, Vladimir-Prelog-Weg 2, 8093 Zurich, Switzerland.}
\author{F. Emil Thomasen}
\email{Authors contributed equally.}
\affiliation{University of Copenhagen, Department of Biology, Linderstr{\o}m-Lang Centre for Protein Science, Ole Maal{\o}es Vej 5, DK-2200, Copenhagen N, Denmark.}
\author{William Bro-J{\o}rgensen}
\email{Authors contributed equally.}
\affiliation{University of Copenhagen, Department of Chemistry and Nano-Science Center, Universitetsparken 5, DK-2100, Copenhagen Ø, Denmark.}
\author{Matthew S. Teynor}
\affiliation{University of Copenhagen, Department of Chemistry and Nano-Science Center, Universitetsparken 5, DK-2100, Copenhagen Ø, Denmark.}
\affiliation{University of Copenhagen, Niels Bohr Institute, NNF Quantum Computing Programme, Blegdamsvej 21, DK-2100 Copenhagen Ø, Denmark.}
\author{Valentina Sora}
\affiliation{University of Copenhagen, Department of Computer Science, Universitetsparken 1, DK-2100, Copenhagen Ø, Denmark.}
\author{Thomas Weymuth}
\affiliation{ETH Zurich, Department of Chemistry and Applied Biosciences, Vladimir-Prelog-Weg 2, 8093 Zurich, Switzerland.}
\author{Raphael T. Husistein}
\affiliation{ETH Zurich, Department of Chemistry and Applied Biosciences, Vladimir-Prelog-Weg 2, 8093 Zurich, Switzerland.}
\author{Frederik E. Knudsen}
\affiliation{University of Copenhagen, Department of Biology, Linderstr{\o}m-Lang Centre for Protein Science, Ole Maal{\o}es Vej 5, DK-2200, Copenhagen N, Denmark.}
\renewcommand*{\thefootnote}{\fnsymbol{footnote}}

\author{Anders Krogh}
\email{akrogh@di.ku.dk}
\affiliation{University of Copenhagen, Department of Computer Science, Universitetsparken 1, DK-2100, Copenhagen Ø, Denmark and Center for Health Data Science, Department of Public Health.}
\author{Kresten Lindorff-Larsen}
\renewcommand*{\thefootnote}{\fnsymbol{footnote}}
\email{lindorff@bio.ku.dk}
\affiliation{University of Copenhagen, Department of Biology, Linderstr{\o}m-Lang Centre for Protein Science, Ole Maal{\o}es Vej 5, DK-2200, Copenhagen N, Denmark.}

\author{Markus Reiher}
\renewcommand*{\thefootnote}{\fnsymbol{footnote}}
\email{mreiher@ethz.ch}
\affiliation{ETH Zurich, Department of Chemistry and Applied Biosciences, Vladimir-Prelog-Weg 2, 8093 Zurich, Switzerland.}

\author{Gemma C. Solomon}
\renewcommand*{\thefootnote}{\fnsymbol{footnote}}
\email{gsolomon@chem.ku.dk}
\affiliation{University of Copenhagen, Department of Chemistry and Nano-Science Center, Universitetsparken 5, DK-2100, Copenhagen Ø, Denmark.}
\affiliation{University of Copenhagen, Niels Bohr Institute, NNF Quantum Computing Programme, Blegdamsvej 21, DK-2100 Copenhagen Ø, Denmark.}

\date{March 05, 2025}

\begin{abstract}

Binding free energies are a key element in understanding and predicting the strength of protein--drug interactions. While classical free energy simulations yield good results for many purely organic ligands, drugs including transition metal atoms often require quantum chemical methods for an accurate description. We propose a general and automated workflow that samples the potential energy surface with hybrid quantum mechanics/molecular mechanics (QM/MM) calculations and trains a machine learning (ML) potential on the QM energies and forces to enable efficient alchemical free energy simulations. To represent systems including many different chemical elements efficiently and to account for the different description of QM and MM atoms, we propose an extension of element-embracing atom-centered symmetry functions for QM/MM data as an ML descriptor. The ML potential approach takes electrostatic embedding and long-range electrostatics into account. We demonstrate the applicability of the workflow on the well-studied protein--ligand complex of myeloid cell leukemia 1 and the inhibitor 19G and on the anti-cancer drug NKP1339 acting on the glucose-regulated protein 78.

\end{abstract}


\maketitle

%%%%%%%%%%%%%%%%%%%%%%%%%%%%%%%%%%%%%%%%%%%%%%%%%%%%%%%%%%%%%%%%%%%%%%%%%%%%%%%%%%%%%%%%%%%%%%%%%%%%%%%%
\section{Introduction}\label{sec:introduction} 
%%%%%%%%%%%%%%%%%%%%%%%%%%%%%%%%%%%%%%%%%%%%%%%%%%%%%%%%%%%%%%%%%%%%%%%%%%%%%%%%%%%%%%%%%%%%%%%%%%%%%%%%

Molecular interactions such as protein-protein, protein-DNA, and protein-RNA interactions are essential for biological functions \cite{Sravan2018}. For example, abnormal protein-protein interactions can cause cancer and neurodegenerative diseases \cite{Lu2020}. Molecular interactions also determine if a drug molecule will bind to a protein target. A key quantity in this binding process is the free energy. If a reduction in free energy is accessible and larger than in competing processes, the drug molecule will preferentially bind to the target protein. This process is called molecular recognition. While there are many molecular mechanisms underlying how drugs assert their functions, such as activating, inhibiting, or targeting a protein for degradation, an essential aspect is binding with sufficient affinity and specificity. Therefore, predicting binding free energies with high accuracy through computational models would allow for targeted and rapid design of new drugs with reduced side effects, reduced risk that bacteria and viruses develop resistance, and reduced economic costs \cite{Mobley2017, Cavasotto2018, Abel2018}. Furthermore, accurate predictive models for molecular interactions can support improved enzyme engineering for technological applications and, more generally, improve understanding of fundamental biological processes \cite{Kumar2006}.

With these considerations in mind, there is clear utility in a generally applicable computational workflow to determine binding free energies from first principles for arbitrary and complex biomolecular systems. Such a workflow should be automated to reduce human time investment and economic cost \cite{DiMasi2016}. While finite computational resources necessitate the introduction of various approximations, the goal is an approach that is systematically improvable with increasing computing power.

Molecular dynamics (MD) simulations can be employed to determine protein--ligand binding free energies. In the simplest approach, the binding free energy can be calculated directly from a standard MD simulation employing the equilibrium populations of the bound and unbound states. However, it can be a major challenge to sample the binding/unbinding process sufficiently. Several different approaches have been developed to solve this sampling problem. One of the most powerful methods is alchemical free energy (AFE) simulations. It bridges the thermodynamic states of interest, e.g., the bound and unbound states, with unphysical (alchemical) intermediate states of the system, circumventing sampling of the binding/unbinding process \cite{Zwanzig1954, Mey2020, Song2020}. Alternatively, a bias potential can be added to overcome the energetic barrier separating the bound and unbound states, such as in metadynamics and umbrella sampling \cite{Torrie1977, Huber1994, Laio2002, Barducci2008, Henin2022}.

Previous studies have shown that free energy calculations with classical force fields can efficiently represent biomolecular systems in solution and may approach experimental accuracy of binding free energies \cite{Tresadern2022, Hahn2024}. However, the applicability of a classical force field is often limited to drug molecules containing main-group elements and those for which experimental data aid in the parameterization of the force field. These limitations can lead to excellent drug candidates being missed in drug discovery attempts,
such as platinum-based drugs, which are well-known chemotherapeutic agents \cite{Zhang2022}. Unfortunately, parametrizing classical force fields for new elements is far from trivial. Moreover, they are challenging to improve systematically \cite{rufa2025fine}.

In contrast to classical force fields, quantum mechanical electronic structure methods can be employed for any element. They typically yield more accurate energies and forces due to their explicit treatment of electrons. However, the increase in accuracy and flexibility also results in significantly higher computational cost. This cost prohibits large-scale sampling of the conformational space of complex molecular systems, which is necessary to estimate the entropic contribution to the free energy accurately. Extensive sampling is especially relevant for biomolecules due to their flexibility, which leads to a large conformational space accessible at physiological temperatures \cite{Huang2012, Chen2013}.

Embedding methods are a powerful tool to circumvent the limitations of both the classical force fields and quantum mechanical electronic structure methods. In a hybrid quantum mechanics/molecular mechanics (QM/MM) approach \cite{Senn2009, Warshel2014}, a small QM region is described by a quantum mechanical electronic structure method, while the remaining atoms are represented by force fields. In this way, the QM/MM approach combines an accurate representation of the region of interest, e.g., a catalytic center in an enzyme or a drug molecule in a protein's binding pocket, with an efficient description of the remaining protein and solvent. We note that another approach to incorporate QM/MM information in a sample-efficient manner is the reformulation of the force field expression to include higher-order physical effects such as multipolar electrostatics and anisotropic polarization \cite{Nawrocki2022}.

Even with the advantages of QM/MM approaches, sampling driven by QM/MM methods will be demanding if the relevant region for the binding process is large, a computationally costly electronic structure method is required, or the sampling procedure converges slowly. Furthermore, AFE approaches employ unphysical intermediate states that cannot easily be described with traditional QM/MM methods. Therefore, QM/MM has previously been applied as an additional correction to an initial binding free energy calculated with MM through QM/MM energy evaluation on MM structures \cite{Beierlein2011, Dybeck2016}. These approaches avoid explicit propagation on the QM/MM potential energy surface (PES). However, they implicitly assume that the MM structures are also sufficiently representative of the structures that would occur in a QM/MM simulation, which may be qualitatively incorrect \cite{CaveAyland2014}.

To increase the sampling efficiency, machine learning (ML) potentials \cite{Behler2007, Behler2021, Boeselt2021, Hofstetter2022} can be applied to enable explicit propagation on the QM/MM PES. ML potentials are able to retain the high accuracy of the PES obtained by a quantum mechanical electronic structure method, while their computational efficiency is comparable to that of a classical force field \cite{Behler2016}. Despite their promising performance for a wide variety of problems and systems \cite{Hellstroem2016, Smith2017, Eckhoff2019, Eckhoff2021a}, the application of ML potentials in the context of protein--drug interactions is still challenging. In hybrid QM/MM approaches, the interactions among QM atoms, among MM atoms, and between QM and MM atoms are treated differently. However, standard ML potentials normally do not differentiate between these interaction types. Furthermore, many structural descriptors applied as features for standard ML potentials cannot deal efficiently with a large number of different chemical elements occurring in protein--drug complexes. Therefore, many previous studies on ML potentials for AFE simulations were limited by the restricted applicability of pre-trained ML potentials, treated the interactions between QM and MM atoms only on the MM level using fixed point charges \cite{Galvelis2023, Zariquiey2024, Grassano2024} or predicted these on the fly during the simulation \cite{Semelak2024}.

Recent work has shown that, for example, a high-dimensional neural network potential (HDNNP) \cite{Behler2007, Behler2017, Behler2021} is able to represent a QM/MM PES including polarization of the QM atoms by MM point charges \cite{Boeselt2021}. However, this approach employs modified atom-centered symmetry functions (ACSFs) as features, which show an unfavorable scaling with the number of different elements. One approach to mitigate this unfavorable scaling is applying element-embracing atom-centered symmetry functions (eeACSFs) \cite{Eckhoff2023}. In this work, we combine these two methods by proposing an extension of eeACSFs, which makes them applicable to QM/MM data. In this way, we can incorporate accurate electronic structure energies obtained in representative chemical environments through efficient ML potentials into large-scale AFE simulations of binding free energies.

In addition to an increase in accuracy, the approach should be integrated into an automated pipeline to decrease human time investment. Therefore, we propose an end-to-end pipeline utilizing distributed computing that begins with system preparation and ends at a binding free energy prediction.

Computationally demanding tasks such as QM/MM calculations can be distributed to multiple computing centers with one central database coordinating the prioritization of tasks and storing results. We apply active learning by a query-by-committee strategy \cite{Artrith2012, Behler2015, Podryabinkin2017, Bernstein2019} to complete the sampling of QM/MM training data in a self-driven way. We then perform AFE simulations and non-equilibrium (NEQ) switching simulations to correct the end state of classical MM simulations.\cite{Rufa2020} These end-state corrections are driven by the hybrid ML/MM approach, which samples the targeted conformation space directly. 
%%%%%%%%%%%%%%%%%%%%%%%%%%%%%%%%%%%%%%%%%%%%%%%%%%%%%%%%%%%%%%%%%%%%%%%%%%%%%%%%%%%%%%%%%%%%%%%%%%%%%%%%
\section{Protein--Ligand Complexes}\label{sec:systems}
%%%%%%%%%%%%%%%%%%%%%%%%%%%%%%%%%%%%%%%%%%%%%%%%%%%%%%%%%%%%%%%%%%%%%%%%%%%%%%%%%%%%%%%%%%%%%%%%%%%%%%%%

\begin{figure}[htb!]
    \centering
    \includegraphics[width=\columnwidth]{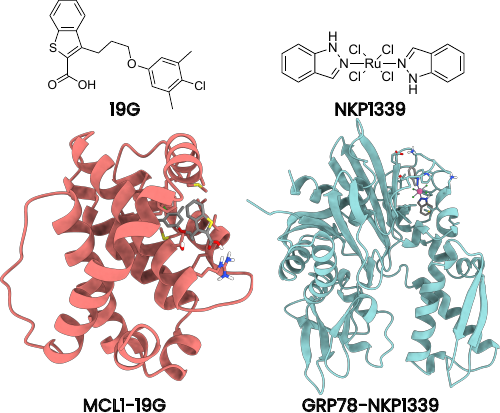}
    \caption{Illustration of the two investigated systems MCL1-19G (left) and GRP78-NKP1339 (right). The ligands are at the top, and the protein--ligand complexes are at the bottom.}
    \label{fig:systems}
\end{figure}

We chose two protein--ligand complexes to test our approach. The first system is myeloid cell leukemia 1 (MCL1) bound to the small molecule 19G \cite{Friberg2012} (see Figure \ref{fig:systems}). MCL1 is a modulator of apoptosis from the BCL2 protein family. Dysregulation of MCL1 is associated with various cancers, and several small molecule inhibitors for MCL1 have been designed for cancer therapy \cite{Wang2021}. MCL1-19G is a suitable model system for our purpose for several reasons: (i) MCL1 is a small protein, (ii) the structure of the complex has been solved by X-ray crystallography \cite{Friberg2012}, (iii) the MCL1-19G affinity has been measured experimentally \cite{Friberg2012}, and (iv) relative AFE calculations have previously been performed for MCL1-19G binding \cite{Wang2015,Hahn2024}.

The second protein--ligand system is the 78 kDa glucose-regulated protein (GRP78) bound to the small molecule NKP1339. GRP78 is an HSP70 chaperone in the endoplasmic reticulum (ER), where it facilitates protein folding, degradation, and translocation across the ER membrane, as well as regulation of the unfolded protein response associated with ER stress \cite{Li2006}. GRP78 also functions as a receptor on the cell surface \cite{Gonzalez2009}. Elevated GRP78 levels and increased localization to the cell membrane are associated with a variety of cancers and the development of chemoresistance \cite{Wang2009,Roller2013}. NKP1339 is a small-molecule anti-cancer compound that targets GRP78 \cite{Trondl2014, Flocke2016}. Phase I clinical trials have shown that NKP1339 has modest anti-tumor activity and a manageable safety profile \cite{Burris2016}. The GRP78--NKP1339 system is a suitable test for our hybrid QM/MM approach because NKP1339 contains a central ruthenium atom, which is typically not parameterized in MM force fields.

%%%%%%%%%%%%%%%%%%%%%%%%%%%%%%%%%%%%%%%%%%%%%%%%%%%%%%%%%%%%%%%%%%%%%%%%%%%%%%%%%%%%%%%%%%%%%%%%%%%%%%%%
\section{Methods}\label{sec:methods}
%%%%%%%%%%%%%%%%%%%%%%%%%%%%%%%%%%%%%%%%%%%%%%%%%%%%%%%%%%%%%%%%%%%%%%%%%%%%%%%%%%%%%%%%%%%%%%%%%%%%%%%%

%%%%%%%%%%%%%%%%%%%%%%%%%%%%%%%%%%%%%%%%%%%%%%%%%%%%%%%%%%%%%%%%%%%%%%%%%%%%%%%%%%%%%%%%%%%%%%%%%%%%%%%%
\subsection{Thermodynamic Cycle} % Moritz
%%%%%%%%%%%%%%%%%%%%%%%%%%%%%%%%%%%%%%%%%%%%%%%%%%%%%%%%%%%%%%%%%%%%%%%%%%%%%%%%%%%%%%%%%%%%%%%%%%%%%%%%

AFE simulations are routinely performed with MM models. These models are ideal as different interactions can be independently tuned; the Lennard-Jones and Coulomb interactions between molecules can gradually be turned off through a series of states defined by a scaling parameter $\lambda$ which scales the non-bonded interactions between specified parts of the system. For example, the ligand can be decoupled from its environment through a series of simulations interpolating between $\lambda$=1, where the system is fully interacting, and $\lambda$=0, where the ligand is in vacuum with the remainder of the system still interacting.

To estimate the free energy differences between the series of states, we utilize the Multistate Bennett acceptance ratio (MBAR) free energy estimator \cite{Shirts2008}. MBAR takes into account the energies of the configurations sampled at every thermodynamic state evaluated with the energy function of every thermodynamic state. This estimation is done by self-consistently solving a system of equations defining the free energy at each thermodynamic state,

\begin{align}
f_i = -\ln \sum^K_{j=1} \sum^{N_j}_{n=1} \frac{\exp[-u_i(x_{jn})]}{\sum^K_{k=1} N_k \exp[f_k - u_k(x_{jn})]}\ ,
\end{align}

where $f_i$ is the free energy of thermodynamic state $i$, $K$ is the total number of thermodynamic states, $N$ is the number of configurations, $u(x)$ is the energy of the configuration $x$ evaluated with the energy function $u$, and $f_k$ is the free energy of another thermodynamic state $k$. The gradual decoupling of interactions using a series of $\lambda$ values is carried out to ensure that there is sufficient overlap between the energy distributions of neighboring states. This approach leads to a more accurate estimate of the free energy differences with MBAR. The overlap can be further improved by applying a Hamiltonian replica exchange scheme, which allows simulations of different thermodynamic states to swap their energy functions on the fly \cite{Woods2003,Mey2020}.

An advantage of MM models for AFE simulations is that they are numerically robust even for unphysical states. These states are encountered if the interaction strength is very low, which can lead to extremely short interatomic distances. To describe these states with ML potentials, the ML potential has to be specifically designed for that purpose \cite{Moore2024}. To avoid this issue, we employ an end-state correction strategy \cite{Hudson2015, Rufa2020, Tkaczyk2024} exploiting that the free energy is a state function. This fact allows us to simulate the alchemical path with the MM force field and then formally switch the PES from MM to ML/MM, correcting for any differences in the PESs.

\begin{figure*}[htb!]
    \centering
    \includegraphics[width=0.666\textwidth]{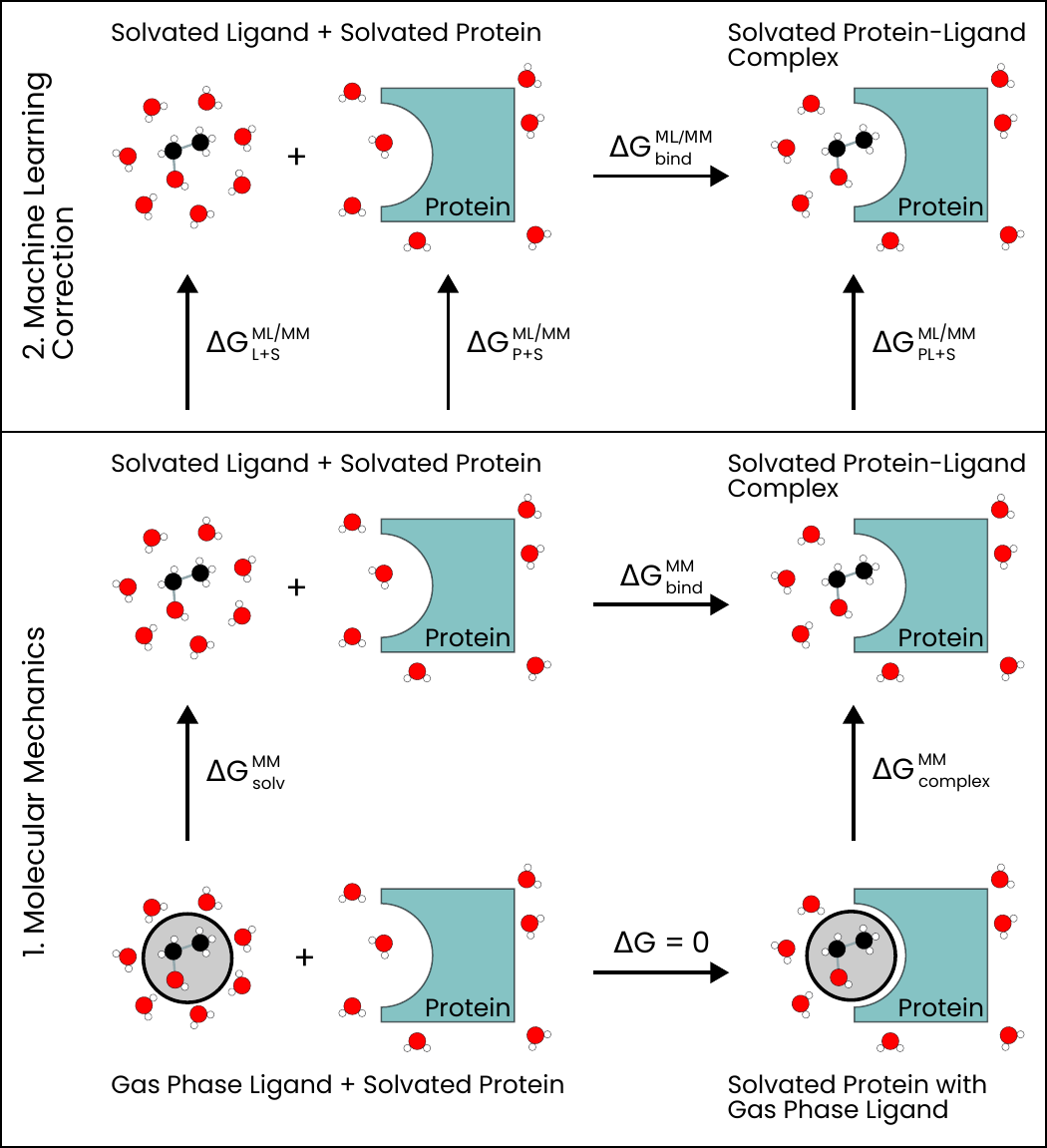}
    \caption{The thermodynamic cycle for calculating the binding free energy with AFE simulations employing MM (bottom box) and ML/MM end-state corrections (top box). The gray-shaded circles represent that the interactions between the ligand and the protein and solvent have been scaled to zero. The white spheres are hydrogen, the red spheres are oxygen, and the black spheres are carbon.}
    \label{fig:thermodynamic-cycle}
\end{figure*}

In AFE simulations, the ML/MM binding free energy of a ligand to a protein $\Delta G_\mathrm{bind}^\mathrm{ML/MM}$ is indirectly calculated as illustrated in the thermodynamic cycle in the bottom box of Figure \ref{fig:thermodynamic-cycle}. Initially, the change in free energy between the solvated protein--ligand complex and the solvated protein with the gas phase ligand is determined in AFE simulations employing an MM force field ($\Delta G_\mathrm{complex}^\mathrm{MM}$ in \autoref{fig:thermodynamic-cycle}). Subsequently, the change in free energy between the solvated ligand and the gas phase ligand is calculated ($\Delta G_\mathrm{solv}^\mathrm{MM}$ in \autoref{fig:thermodynamic-cycle}). The change in free energy between the two states with the ligand in gas phase is zero ($\Delta G = 0$). Therefore, the free energy of binding can be determined as

\begin{align}
    \Delta G_\mathrm{bind}^\mathrm{MM} = \Delta G_\mathrm{complex}^\mathrm{MM} - \Delta G_\mathrm{solv}^\mathrm{MM}\ .
    \label{eq:total-free-energy-of-binding-MM}
\end{align}

Next, we calculate the correction to the free energy associated with going from the MM description to the ML/MM description for the end states in the thermodynamic cycle (top box of \autoref{fig:thermodynamic-cycle}). We apply NEQ switching simulations to determine the free energy difference between MM and ML/MM for the protein--ligand complex $\Delta G_\mathrm{PL+S}^\mathrm{ML/MM}$ and for the ligand in solution $\Delta G_\mathrm{L+S}^\mathrm{ML/MM}$.

There is no need to calculate an ML/MM correction for the gas-phase ligand as this correction does not contribute to the thermodynamic cycle. Furthermore, the ML/MM correction for the solvent $\Delta G_\mathrm{S}^\mathrm{ML/MM}$ and the protein in solution $\Delta G_\mathrm{P+S}^\mathrm{ML/MM}$ are zero in our approach because we restrict the ML potential to the ligand. This restriction is reasonable since the protein and solvent can be described by specialized and accurate MM force fields, which are not available for arbitrary ligands.

Adding all corrections to Equation (\ref{eq:total-free-energy-of-binding-MM}), we obtain
\begin{align}
    \begin{split}
        &\Delta G_\mathrm{bind}^\mathrm{ML/MM} = \Delta G_\mathrm{complex}^\mathrm{MM}
        + \Delta G_\mathrm{PL+S}^\mathrm{ML/MM}\\
        &- \Delta G_\mathrm{solv}^\mathrm{MM}
        % - \Delta G_\mathrm{P+S}^\mathrm{ML/MM}
        - \Delta G_\mathrm{L+S}^\mathrm{ML/MM}~.
    \label{eq:total-free-energy-of-binding-1}
    \end{split}
\end{align}

%%%%%%%%%%%%%%%%%%%%%%%%%%%%%%%%%%%%%%%%%%%%%%%%%%%%%%%%%%%%%%%%%%%%%%%%%%%%%%%%%%%%%%%%%%%%%%%%%%%%%%%%
\subsection{Workflow} 
%%%%%%%%%%%%%%%%%%%%%%%%%%%%%%%%%%%%%%%%%%%%%%%%%%%%%%%%%%%%%%%%%%%%%%%%%%%%%%%%%%%%%%%%%%%%%%%%%%%%%%%%

\begin{figure}[htb!]
    \centering
    
    \includegraphics[width=\columnwidth]{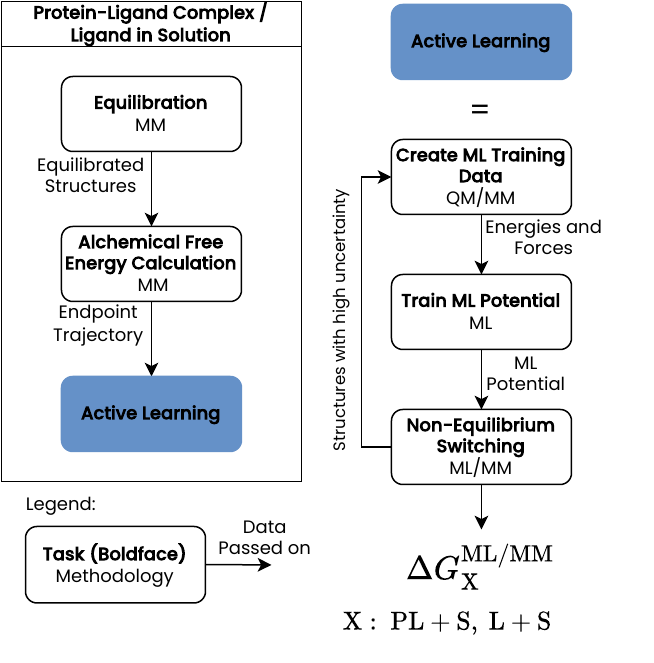}
    \caption{
    Workflow to determine the free energy of binding by ML/MM starting from a protein-ligand complex structure and a solvated ligand structure.
    }
    %%%%%%%%%%%%%%%%%%%%%%%%%%%%%%%%%%%%%%%%%%%%%%%%%%%%%
    \label{fig:workflow}
\end{figure}

As established in Section 3.1, we need to calculate $\Delta G_\mathrm{complex}^\mathrm{MM}$ and $\Delta G_\mathrm{solv}^\mathrm{MM}$ with MM-driven AFE simulations, as well as the end-state correction terms for the protein--ligand complex $\Delta G_\mathrm{PL+S}^\mathrm{ML/MM}$

and the ligand in solution $\Delta G_\mathrm{L+S}^\mathrm{ML/MM}$ by NEQ switching from the MM PES to the ML/MM PES. Our workflow to calculate these terms is shown in Figure \ref{fig:workflow}.

The first step is equilibrating the protein--ligand complex since we must perform AFE simulations, starting from the equalibrated system, to calculate the work required to transfer the ligand from the protein to vacuum and then solvate the ligand.

The initial training data set of the ML potential is based on structures from MD trajectories of both end states, the fully interacting ligand--protein complex and the solvated ligand. The chosen structures need to be well separated in time to reduce their correlation. QM/MM energies and forces are calculated as target properties for the initial training of the ML potential. Subsequently, NEQ switching simulations are carried out employing the ML potential to estimate the PES switching corrections $\Delta G_\mathrm{PL+S}^\mathrm{ML/MM}$

and $\Delta G_\mathrm{L+S}^\mathrm{ML/MM}$. Any structure encountered during these simulations that shows a high uncertainty in the energy or force predictions is collected. A selection of these structures, with a minimum distance in time to avoid too strong correlation, is recalculated employing QM/MM and retrained by the ML potential. 

%%%%%%%%%%%%%%%%%%%%%%%%%%%%%%%%%%%%%%%%%%%%%%%%%%%%%%%%%%%%%%%%%%%%%%%%%%%%%%%%%%%%%%%%%%%%%%%%%%%%%%%%
\subsection{Quantum Mechanics in Molecular Mechanics Embedding\label{sec:QMMM}} % Moritz
%%%%%%%%%%%%%%%%%%%%%%%%%%%%%%%%%%%%%%%%%%%%%%%%%%%%%%%%%%%%%%%%%%%%%%%%%%%%%%%%%%%%%%%%%%%%%%%%%%%%%%%%

In the hybrid QM/MM approach, the full system containing protein, ligand, and solvent (including solvated ions) is partitioned into a QM region $Q$ and the remaining MM environment $E$. The interaction between both regions can be described by molecular mechanics or in a more elaborate approach using electrostatic embedding. We employ the latter approach, in which the electronic wave function of the QM region is polarized by the Coulomb potential of the MM point charges. The QM/MM energy is
\begin{align}
    E_\mathrm{QM/MM} = E_\mathrm{QM} + E^\mathrm{int}_\mathrm{elec} + E^\mathrm{int}_\mathrm{MM} + E_\mathrm{MM}\ ,\label{eq:E_QMMM}
\end{align}
where $E_\mathrm{QM}$ is the energy of the QM region, $E^\mathrm{int}_\mathrm{elec}$ is the electrostatic interaction between QM and MM region, $E^\mathrm{int}_\mathrm{MM}$ collects all non-electrostatic interactions, and $E_\mathrm{MM}$ is the energy of the MM region.

The electrostatic interaction $E^\mathrm{int}_\mathrm{elec}$ is given as
\begin{align}
    \begin{split}
        E^\mathrm{int}_\mathrm{elec} &= \left\langle \Psi_{Q}\left| \sum_{A\in E} \frac{q_A}{\left|\pmb{r} - \pmb{R}_A\right|} \right| \Psi_{Q}\right\rangle\\
        &+ \sum_{I\in {Q}} \sum_{A\in E} \frac{Z_I q_A}{\left| \pmb{R}_I - \pmb{R}_A \right|}\ .
    \label{eq:electrostatic_embedding}        
    \end{split}
\end{align}
Here, $\Psi_{Q}$ denotes the wave function of the QM region. $q_A$ is a point charge with MM index $A$ at position $\pmb{R}_A$ representing the electrostatic potential of the MM environment. $Z_I$ is the nuclear charge of QM atom $I$ at position $\pmb{R}_I$ in the QM region.

$E^\mathrm{int}_\mathrm{MM}$ includes all non-electrostatic interactions between QM and MM region which are calculated by the MM force field.

%%%%%%%%%%%%%%%%%%%%%%%%%%%%%%%%%%%%%%%%%%%%%%%%%%%%%%%%%%%%%%%%%%%%%%%%%%%%%%%%%%%%%%%%%%%%%%%%%%%%%%%%
\subsection{Machine Learning Potentials} \label{sec:mlp}% Marco
%%%%%%%%%%%%%%%%%%%%%%%%%%%%%%%%%%%%%%%%%%%%%%%%%%%%%%%%%%%%%%%%%%%%%%%%%%%%%%%%%%%%%%%%%%%%%%%%%%%%%%%%
We choose second-generation HDNNP \cite{Behler2007, Behler2017, Behler2021} where the energy of a system with $N_\mathrm{elem}$ chemical elements and $N_\mathrm{atom}^m$ atoms $n$ of element $m$ is constructed as a sum of atomic energy contributions $E_{\mathrm{atom},n}^m$. Feed-forward neural networks are employed to represent these atomic energy contributions and, in this work, the neural network consists of two hidden layers with a linear output \cite{Eckhoff2023}. The neural network input is a vector $\mathbf{G}_n^m$ of dimension $n_G$, which describes the local atomic environment of atom $n$.

\subsubsection*{Element-Embracing Atom-Centered Symmetry Functions}

Originally, the structural descriptor of HDNNPs is a vector of ACSFs \cite{Behler2011}. ACSFs are many-body representations of the interatomic distances and angles within a cutoff sphere. They fulfill the translational, rotational, and permutational invariances of the potential energy surface. Moreover, the ACSF vector size does not change with the atomic environment making ACSF vectors applicable as input of pre-trained feed-forward neural networks. Further, ACSFs enable the simulation of chemical reactions since they do not employ connectivities. However, each ACSF value represents only the distances or angles for a certain chemical element pair or triple. The scaling in the vector size is, therefore, unfavorable with respect to the number of chemical elements.

Element-embracing atom-centered symmetry functions (eeACSFs) \cite{Eckhoff2023} solve this issue by including an explicit dependence on element information from the periodic table by the term $H_{i,j}^\mathrm{rad}$ and $H_{i,jk}^\mathrm{ang}$ in the radial eeACSFs,
\begin{align}
G_{n,i}^{\mathrm{rad}}=\sum_{j\neq n}^{N_\mathrm{atom}}I_{i,j}^\mathrm{rad}\cdot H_{i,j}^\mathrm{rad}\cdot F_{i,j}^\mathrm{rad}\left(R_{nj}\right)\ ,
\label{eq:radial_functions}
\end{align}
and angular eeACSFs,
\begin{align}
\begin{split}
&G_{n,i}^{\mathrm{ang}}=\sum_{j\neq n}^{N_\mathrm{atom}}\sum_{k<j\land k\neq n}^{N_\mathrm{atom}}I_{i,jk}^\mathrm{ang}\cdot H_{i,jk}^\mathrm{ang}\cdot F_{i,jk}^\mathrm{ang}\left(\theta_{njk}\right)\\
&\cdot F_{i,j}^\mathrm{rad}\left(R_{nj}\right)\cdot F_{i,k}^\mathrm{rad}\left(R_{nk}\right)\ ,
\end{split}
\label{eq:angular_functions}
\end{align}
respectively. The additional interaction type dependent functions $I_{i,j}^\mathrm{rad}$ and $I_{i,jk}^\mathrm{ang}$ will be introduced in the next subsection. In this work, the radial structure is described by

\begin{align}
\begin{split}
&F_{i,j}^\mathrm{rad}\left(R_{nj}\right)\\
&=\begin{cases}\exp\left\{\eta_i\left[1-\left(1-\frac{R_{nj}^2}{R_\mathrm{c}^2}\right)^{-1}\right]\right\}&\mathrm{for}\ R_{nj}<R_\mathrm{c}\\
0&\mathrm{otherwise}\end{cases}\ ,    
\end{split}
\end{align}
which depends on the distances $R_{nj}$ between the central atom $n$ and the neighbor $j$. This function also damps the eeACSF value and all its derivatives smoothly to zero at the cutoff radius $R_\mathrm{c}$. We note that in the original ACSF and eeACSF approach, a separate cutoff function was applied on top of the radial function \cite{Behler2011, Eckhoff2023}. For the angular structure representation, we apply the conventional function
\begin{align}
F_{i,jk}^\mathrm{ang}\left(\theta_{njk}\right)=\left[\frac{1}{2}+\frac{\lambda_i}{2}\cos\left(\theta_{njk}\right)\right]^{\zeta_i}\ ,
\end{align}
with the angles $\theta_{njk}$ between atom $n$ and the neighbors $j$ and $k$. The application of various parameter values for $\eta_i>0$, $\lambda_i=\pm1$, and $\zeta_i\geq1$ in each eeACSF $i$ can yield a structural fingerprint of atom $n$'s local atomic environment.

The initial approach for explicit element-dependence in ACSFs is given by weighted atom-centered symmetry functions \cite{Gastegger2018}. This descriptor employs the atomic numbers of the neighbor atoms to differentiate among various elements. To obtain a more balanced description of light and heavy elements and to exploit trends of the periodic table, the eeACSF vector includes different element properties $h$ for the element-dependent term $H$,
\begin{align}
h_{i,j}\in\left\{1,\ n_j,\ m_j,\ d_j,\ \overline{n}_j,\ \overline{m}_j,\ \overline{d}_j\right\}\ .
\end{align}
These properties are the element's period number in the periodic table $n$, the group number in the s- and p-block $m$ (main group 1 to 8), and the group number in the d-block $d$ ($d=0$ for main group elements). For a balanced representation of all elements, these numbers are also counted in reverse order in the properties $\overline{n}_j\coloneqq X-n_j$, $\overline{m}_j\coloneqq9-m_j$, and $\overline{d}_j\coloneqq11-d_j$ ($\overline{d}=0$ for main group elements). In addition, element-independent eeACSF values are included in the eeACSF vector by setting $h_{i,j}=1$. We note that extensions to the f-block can follow the scheme of the d-block. In this work, we apply the maximum period number $X-1=5$, i.e., the heaviest element can be xenon. For d-block elements, we apply $m=2$ and $\overline{m}=7$.

In the radial eeACSFs these element properties are divided by their maximal possible value $h_i^\mathrm{max}$ of the element property $h$ utilized in eeACSF $i$, to keep the element-dependent terms of each neighbor atom $j$ between 0 and 1,
\begin{align}
H_{i,j}^\mathrm{rad}=\frac{h_{i,j}}{h_i^\mathrm{max}}\ .\label{eq:H_rad}
\end{align}
The angular eeACSFs employ linear combinations of the element properties of neighbor $j$ and $k$ with hyperparameter $\gamma_i=\pm1$,
\begin{align}
H_{i,jk}^\mathrm{ang}=\frac{\left|h_{i,j}+\gamma_ih_{i,k}\right|+\frac{1-\gamma_i}{2}C_{ijk}}{h_i^\mathrm{max}\left(\frac{1+\gamma_i}{2}+1\right)}\ ,
\end{align}
with 
\begin{align}
C_{ijk}=\begin{cases}0&\mathrm{for}\ h_{i,j}=h_{i,k}=0\\
1&\mathrm{otherwise}\end{cases}\ .
\end{align}
The value of the absolute linear combination is shifted by one if the difference ($\gamma_i=-1$) is applied. This approach ensures that each contribution is taken into account. If the element properties of both neighbors are equal to zero, this shift is not applied. The resulting value is divided by $2h_i^\mathrm{max}$ for sums ($\gamma_i=1$) and by $h_i^\mathrm{max}$ for differences ($\gamma_i=-1$). In the end, the eeACSF vector is computationally more efficient than the ACSF vector when about four or more different chemical elements need to be represented.

\subsubsection*{Representation of QM/MM Interactions in Element-Embracing Atom-Centered Symmetry Functions}

ML potentials are constructed to obtain energies and atomic forces with first principles quality but at an efficiency close to MM approaches. The most reasonable strategy to speed up QM/MM calculations with electrostatic embedding is, therefore, to learn the QM-related energy contribution $E_\mathrm{QM}+E^\mathrm{int}_\mathrm{elec}$ and to evaluate $E^\mathrm{int}_\mathrm{MM}+E_\mathrm{MM}$ by the MM approach (see Equation (\ref{eq:E_QMMM})). Since the ML potential descriptor takes into account only the local atomic environment, the long-range electrostatic interactions should be handled differently. Therefore, the electrostatic interactions between mixed QM--MM atom pairs are subtracted from the energy to be learned,
\begin{align}
E_\mathrm{ML}=E_\mathrm{QM}+E^\mathrm{int}_\mathrm{elec}-\sum_{I\in Q}\sum_{A\in E}\frac{q_I q_A}{|\pmb{R}_I - \pmb{R}_A|}\ .
\end{align}
In this work, the atomic charge of QM atom $I$ is taken from the MM force field parameters. Therefore, a difference to the actual electrostatic interaction $E^\mathrm{int}_\mathrm{elec}$ exists, which needs to be covered by the ML potential. However, if the MM and QM charge distributions are similar, the difference should be small for large distances and the ML potential is able to account for the interaction within the cutoff radius. A more elaborate approach would be to learn environment-dependent atomic charges obtained from the QM calculations in a second machine learning model. Subsequently, these charges can be applied to calculate the long-range electrostatic interactions as in third-generation HDNNPs \cite{Artrith2011}. We note that the MM atoms within the cutoff radius of a QM atom (environment region $E^\prime$) contribute to the ML potential energy and hence their forces have MM and ML potential contributions.

In addition, we note that competing approaches are trained on the difference between QM data and data from a semi-empirical method \cite{Boeselt2021, Hofstetter2022}. This so-called delta learning can increase the accuracy, but it has the drawback of requiring semi-empirical method calculations for inference. To maximize efficiency, we avoid delta learning in our approach.

The energy to be learned still depends on the positions of QM and MM atoms. However, the eeACSF vector is constructed to represent atomic environments in which all atoms are treated employing the same approach. For ACSFs this limitation was circumvented by introducing a new element type for MM atoms and multiplying the ACSF contributions by the respective atomic charges of the neighbors \cite{Boeselt2021}. To overcome this limitation also for eeACSFs, we propose an interaction-dependent term $I$ in the eeACSFs, which is inspired by the spin-dependent ACSF approach \cite{Eckhoff2021}. The central atom $n$ of an eeACSFs can always be chosen to be a QM atom since we aim to calculate its energy contribution and electrostatic interaction with the MM atoms. To differentiate the interactions from a QM atom to its QM neighbors $Q$ and MM neighbors $E$, we introduce the atom type properties $t^Q$ and $t^E$. The former is one for QM neighbors and zero for MM neighbors, and the latter is vice versa. In this way, we can construct radial eeACSFs describing interactions only to QM atoms and only to MM atoms by the interaction-dependent term
\begin{align}
I_{i,j}^\mathrm{rad}\in\left\{t_j^Q,\ t_j^E\right\}\ .
\end{align}
The angular eeACSFs differentiate between interactions with two QM neighbors, one QM and one MM neighbor, and two MM neighbors,
\begin{align}
I_{i,jk}^\mathrm{ang}\in\left\{t_j^Q\cdot t_k^Q,\ t_j^Q\cdot t_k^E+t_j^E\cdot t_k^Q,\ t_j^E\cdot t_k^E\right\}\ .
\end{align}
Analogous to the element-dependent terms, different entries in the eeACSF vector can employ different interaction-dependent terms to represent all interactions.

The MM atoms enter the energy $E_\mathrm{QM}+E^\mathrm{int}_\mathrm{elec}$ by their electrostatic interaction with the QM atoms. Therefore, instead of the chemical element, the atomic charge is the only representative property of these atoms besides their positions. As a consequence, the element-dependent term in the radial eeACSF (Equation (\ref{eq:H_rad})) is replaced by 
\begin{align}
H_{i,j_E}^\mathrm{rad}=\frac{q_{j_E}}{2\,q_\mathrm{max}}\ ,
\end{align}
if neighbor $j$ is an MM atom. Its atomic charge $q_{j_E}$ is divided by two times the hyperparameter of the maximal possible absolute atomic charge $q_\mathrm{max}$. In this way, the maximal range of the $H$ values is similar to those of interactions to QM neighbors, while larger absolute values of the atomic charges still lead to stronger interactions. For the angular eeACSFs, products of the terms from neighbors $j$ and $k$ are applied. This ansatz leads to
\begin{align}
H_{i,jk_E}^\mathrm{ang}=\frac{h_{i,j}q_{k_E}}{2\,h_i^\mathrm{max}q_\mathrm{max}}\ ,
\end{align}
if one neighbor is a QM atom and one is an MM atom (indices $j$ and $k$ can be interchanged), and to
\begin{align}
H_{i,j_Ek_E}^\mathrm{ang}=\frac{q_{j_E}q_{k_E}}{2\,q_\mathrm{max}^2}\ ,
\end{align}
if both neighbors are MM atoms.

\subsubsection*{Uncertainty Quantification and Active Learning}

The uncertainty of ML potentials has multiple origins such as from training data and model variance \cite{Heid2023}, and multiple approaches exist to reduce them.\cite{Heid2023} Uncertainty originating from insufficient training data can be alleviated with active learning strategies (explained below). An ensemble (or committee) of ML potentials that are trained on the same training data, but with differently initialized parameters can reduce model variance and estimate the size of this uncertainty \cite{Peterson2017, Smith2018, Musil2019, Imbalzano2021}. Large model variance correlates with training data that is too sparse. The reason is that predictions are arbitrary for unknown structures due to the high flexibility of the machine learning model. In this way, missing training data can be identified, recalculated by the reference method, and retrained by the ML potential. This process is called active learning \cite{Artrith2012, Behler2015, Podryabinkin2017, Bernstein2019}. The retraining of the ML potential during active learning can be accelerated by continual learning strategies. This concept of so-called lifelong ML potentials \cite{Eckhoff2023} exploits algorithms such as lifelong adaptive data selection and the continual resilient (CoRe) optimizer \cite{Eckhoff2023, Eckhoff2024} that enable efficient learning from new data and retaining of previous knowledge. In this way, the ML potential does not have to be retrained from scratch, and only a fraction of the previous training data needs to be employed to mitigate forgetting.

% ensemble model
The energy of the ensemble model $\overline{E}$ is given by the mean of the individual energies $E_p$ of the $N_\mathrm{MLP}$ ML potentials $p$. The energy uncertainty $\Delta\overline{E}$ can be estimated by the sample standard deviation,
\begin{align}
\begin{split}
&\Delta\overline{E}=\mathrm{max}\left\{\overline{\mathrm{RMSE}(E^\mathrm{test})},\right.\\
&\left.\left[\frac{c^2}{N_\mathrm{MLP}-1}\sum_{p=1}^{N_\mathrm{MLP}}\left(E_p-\overline{E}\right)^2\right]^{\tfrac{1}{2}}\right\}\ .
\end{split}
\end{align}
The scaling factor $c$ can calibrate the uncertainty for a certain confidence interval. The minimum uncertainty is set to the mean of the root mean squared errors (RMSEs) of each ML potential's test data. Analogous equations hold for the atomic force components.

%%%%%%%%%%%%%%%%%%%%%%%%%%%%%%%%%%%%%%%%%%%%%%%%%%%%%%%%%%%%%%%%%%%%%%%%%%%%%%%%%%%%%%%%%%%%%%%%%%%%%%%%
\section{Results and Discussion}\label{sec:results}
%%%%%%%%%%%%%%%%%%%%%%%%%%%%%%%%%%%%%%%%%%%%%%%%%%%%%%%%%%%%%%%%%%%%%%%%%%%%%%%%%%%%%%%%%%%%%%%%%%%%%%%%

%%%%%%%%%%%%%%%%%%%%%%%%%%%%%%%%%%%%%%%%%%%%%%%%%%%%%%%%%%%%%%%%%%%%%%%%%%%%%%%%%%%%%%%%%%%%%%%%%%%%%%%%
\subsection{Relative Energy Distributions}
%%%%%%%%%%%%%%%%%%%%%%%%%%%%%%%%%%%%%%%%%%%%%%%%%%%%%%%%%%%%%%%%%%%%%%%%%%%%%%%%%%%%%%%%%%%%%%%%%%%%%%%%

\begin{figure}[htb!]
    \centering
    \includegraphics[width=\columnwidth]{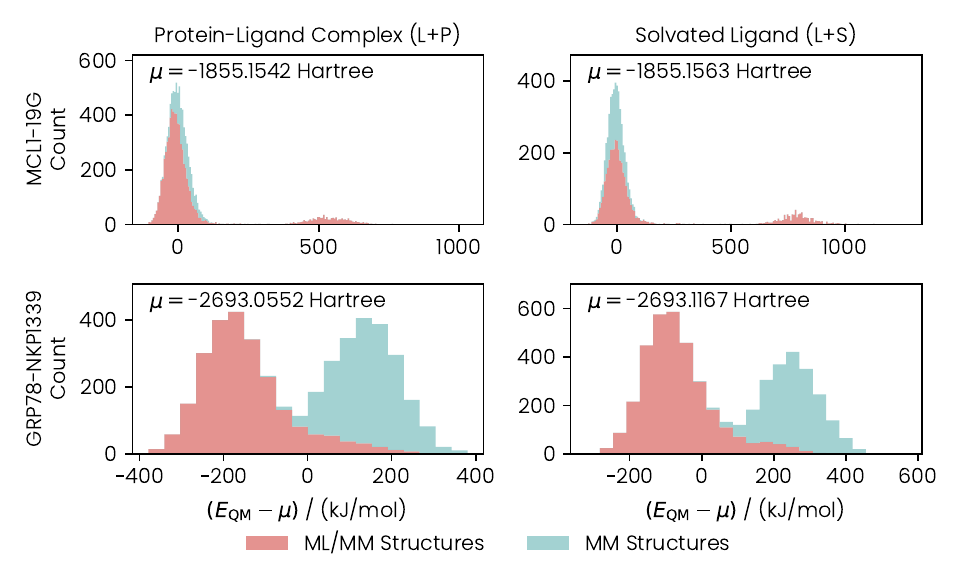}
    \caption{Energy distributions for the QM region calculated for the structures extracted from the end states of the classical AFE simulations and active learning. The distributions were shifted by their median $\mu$. The distributions for the structures from the MM trajectories are stacked on top of the distribution for the structures from active learning.}
    \label{fig:QM-region-energies}
\end{figure}

The target quantity for training the ML potential is the energy of the QM region shifted by the electrostatic interaction between MM and QM region. To illustrate the QM energy values encountered during training, we plotted the distributions of $E_\mathrm{QM}$ in Figure~\ref{fig:QM-region-energies}. In these plots, we color-coded the energies calculated for structures from the end states of the classical AFE simulation (``MM Structures'') and the structures from active learning (``ML/MM Structures'').

% MCL1-19G
The $E_\mathrm{QM}$ distributions for both MCL1-19G end states (protein-ligand complex and the solvated ligand) consist of a large Gaussian close to the median ($0~\si{kJ.mol^{-1}}$ in the plot) and a small peak at high energies. The high-energy structures were generated during the first iteration of active learning and represent transition-state-like structures, as shown in the Supporting Information (Figure S1).
For the solvated ligand, the energies of the Gaussian close to the median are distributed over the same interval independent of whether they were calculated for structures extracted from the MM simulation or obtained during active learning. By contrast, we see a slight shift to higher energies for the MM structures of the protein-ligand complex. In general, we would expect that structures extracted from the active learning are likely to show lower energies than the structures from the MM trajectory because they were obtained during the ML/MM propagation, which allows for structural relaxation of the system according to the ML potential trained on the QM energies. Of course, this assumption does not apply to the first iterations of active learning, where the ML potential is highly uncertain and may produce transition-state-like structures.

% GRP78-NKP1339
Such transition-state-like structures are not encountered for the GRP78-NKP1339 protein-ligand complex and the NKP1339 solvated ligand. However,
the shift between energy distributions from MM and active learning structures is very large for GRP78-NKP1339, as shown in Figure~\ref{fig:QM-region-energies}. For both end states, the distributions for the MM and active learning structures appear as Gaussian distributions, showing only a small overlap. The small overlap between the distributions suggests that the MM force field for GRP78-NKP1339 provides significantly different structures than sampled during the ML/MM propagation. This difference is likely caused by the imperfect MM force field parameters we generated for the Ru center in NKP1339. By contrast, the 19G ligand is purely organic and parameterized with the well-established GAFF force field in the MM simulations.

%%%%%%%%%%%%%%%%%%%%%%%%%%%%%%%%%%%%%%%%%%%%%%%%%%%%%%%%
%%%% Update once the large QM region data is available
%%%%%%%%%%%%%%%%%%%%%%%%%%%%%%%%%%%%%%%%%%%%%%%%%%%%%%%%

%%%%%%%%%%%%%%%%%%%%%%%%%%%%%%%%%%%%%%%%%%%%%%%%%%%%%%%%%%%%%%%%%%%%%%%%%%%%%%%%%%%%%%%%%%%%%%%%%%%%%%%%
\subsection{Accuracy of the Machine Learning Potential}
%%%%%%%%%%%%%%%%%%%%%%%%%%%%%%%%%%%%%%%%%%%%%%%%%%%%%%%%%%%%%%%%%%%%%%%%%%%%%%%%%%%%%%%%%%%%%%%%%%%%%%%%

% active learning
For each system MCL1-19G, 19G, GRP78-NKP1339, and NKP1339 a separate HDNNP ensemble was trained. This approach of several smaller expert models instead of one larger general model leads to more flexibility in the learning process and potentially a higher accuracy-cost ratio in predictions. Each initial training was carried out on about $2\cdot10^3$ training conformations from MM AFE simulations. Active learning, i.e., AFE simulations with current ML potentials, extraction and recalculation of high uncertainty structures, was applied to complete the sampling. Besides the reduction of high uncertainty predictions, the resulting binding free energy was monitored during active learning (Figure S3 in the Supporting Information). With increasing reliability of the ML potentials, the binding free energy converged at a certain value. The final number of reference conformations $N_\mathrm{conf}$, i.e. both training and test conformations, are 9224 for MCL1-19G, 5967 for 19G, 4348 for GRP78-NKP1339, and 5234 for NKP1339, whereby each conformation includes $N_Q=44$ QM atoms for MCL1-19G and 19G and $N_Q=35$ QM atoms for GRP78-NKP1339 and NKP1339.

If we compare the ranges of the reference data (Table \ref{tab:MLP_reference_data}) with the respective RMSEs of the prediction (Table \ref{tab:MLP_accuracy}), we observe that the errors are two orders of magnitude smaller than the values. This trend is true for all three properties: QM-related energies $E_\mathrm{ML}$; QM-related atomic force components of QM atoms $F_{\alpha,n(Q)}$; and MM atoms represented by the ML potential $F_{\alpha,n(E^\prime)}$.

\begin{table*}[htb!]
\caption{Ranges and standard deviations of $E_\mathrm{ML}^\mathrm{ref}$, $F_{\alpha,n(Q)}^\mathrm{ref}$, and $F_{\alpha,n(E^\prime)^\mathrm{ref}}$.}
\begin{center}
\begin{tabular}{lrrrr}
	Reference data                                                                                   & MCL1-19G &      19G & GRP78-NKP1339 & NKP1339\vspace{0.075cm} \\ \toprule
	$\mathrm{range}(E_\mathrm{ML}^\mathrm{ref}/N_Q)\,/\,\mathrm{kJ\,mol}^{-1}$                       & $26.916$ & $31.771$ &      $16.478$ &                $17.868$ \\
	$\mathrm{std}(E_\mathrm{ML}^\mathrm{ref}/N_Q)\,/\,\mathrm{kJ\,mol}^{-1}$                         &  $3.637$ &  $5.872$ &       $4.606$ &                 $4.661$ \\
	$\mathrm{range}(F_{\alpha,n(Q)}^\mathrm{ref})\,/\,\mathrm{kJ\,mol}^{-1}\,\text{\AA}^{-1}$        & $1881.3$ & $1787.0$ &      $1877.5$ &                $1760.7$ \\
	$\mathrm{std}(F_{\alpha,n(Q)}^\mathrm{ref})\,/\,\mathrm{kJ\,mol}^{-1}\,\text{\AA}^{-1}$          &   $86.8$ &   $95.0$ &       $134.9$ &                 $129.5$ \\
	$\mathrm{range}(F_{\alpha,n(E^\prime)}^\mathrm{ref})\,/\,\mathrm{kJ\,mol}^{-1}\,\text{\AA}^{-1}$ & $146.12$ & $136.95$ &      $125.46$ &                $110.91$ \\
	$\mathrm{std}(F_{\alpha,n(E^\prime)}^\mathrm{ref})\,/\,\mathrm{kJ\,mol}^{-1}\,\text{\AA}^{-1}$   &   $1.84$ &   $2.72$ &        $1.88$ &                  $2.61$ \\ \bottomrule
\end{tabular}
\end{center}
\label{tab:MLP_reference_data}
\end{table*}

\begin{table*}[htb!]
	\begin{threeparttable}
		\caption{RMSEs of individual HDNNPs and ensembles for $E_\mathrm{ML}$, $F_{\alpha,n(Q)}$, and $F_{\alpha,n(E^\prime)}$.}
		\begin{center}
		\begin{tabular}{lrrrr}
			%\hline\vspace{-0.325cm}
			%\hline\vspace{-0.325cm}                                                               &  \\
			Individual HDNNP RMSEs                                                                 &        MCL1-19G &             19G &   GRP78-NKP1339 &       NKP1339\vspace{0.075cm} \\ \toprule
			%\hline\vspace{-0.325cm}                                                               &  \\
			$E_\mathrm{ML}^\mathrm{train}N_Q^{-1}\,/\,\mathrm{kJ\,mol}^{-1}$                       & $0.190\pm0.005$ & $0.282\pm0.012$ & $0.181\pm0.008$ &               $0.182\pm0.007$ \\
			$E_\mathrm{ML}^\mathrm{test}N_Q^{-1}\,/\,\mathrm{kJ\,mol}^{-1}$                        & $0.186\pm0.006$ & $0.292\pm0.012$ & $0.179\pm0.008$ &               $0.180\pm0.010$ \\
			$F_{\alpha,n(Q)}^\mathrm{train}\,/\,\mathrm{kJ\,mol}^{-1}\,\text{\AA}^{-1}$            &    $12.3\pm0.1$ &    $15.3\pm0.2$ &    $10.3\pm0.2$ &                   $10.5\pm0.2$ \\
			$F_{\alpha,n(Q)}^\mathrm{test}\,/\,\mathrm{kJ\,mol}^{-1}\,\text{\AA}^{-1}$             &    $11.9\pm0.1$ &    $15.1\pm0.1$ &    $10.0\pm0.1$ &                   $10.2\pm0.2$ \\
			$F_{\alpha,n(E^\prime)}^\mathrm{train}\,/\,\mathrm{kJ\,mol}^{-1}\,\text{\AA}^{-1}$     &   $1.01\pm0.02$ &   $1.78\pm0.04$ &   $1.08\pm0.02$ &                 $1.56\pm0.01$ \\
			$F_{\alpha,n(E^\prime)}^\mathrm{test}\,/\,\mathrm{kJ\,mol}^{-1}\,\text{\AA}^{-1}$      &   $1.00\pm0.01$ &   $1.76\pm0.04$ &   $1.07\pm0.02$ &   $1.55\pm0.01$ \\ \bottomrule
			Ensemble RMSEs\vspace{0.075cm}                                                         &  \\ \toprule
			$\overline{E}_\mathrm{ML}N_Q^{-1}\,/\,\mathrm{kJ\,mol}^{-1}$                           &         $0.158$ &         $0.237$ &         $0.152$ &                       $0.157$ \\
			$\overline{F}_{\alpha,n(Q)}\,/\,\mathrm{kJ\,mol}^{-1}\,\text{\AA}^{-1}$                &          $10.57$ &         $13.20$ &          $9.01$ &                         $9.214$ \\
			$\overline{F}_{\alpha,n(E^\prime)}\,/\,\mathrm{kJ\,mol}^{-1}\,\text{\AA}^{-1}$         &          $0.94$ &          $1.63$ &          $1.03$ &                        $1.51$ \\
			$\overline{F}_{\alpha,n(E^{\prime\prime})}\,/\,\mathrm{kJ\,mol}^{-1}\,\text{\AA}^{-1}$ &          $0.41$ &          $0.63$ &          $0.45$ &                        $0.71$ \\ \bottomrule
		\end{tabular}
		\begin{tablenotes}
			\footnotesize
			\item The individual HDNNPs were evaluated on the final training data, i.e., without data declared to be redundant or inconsistent, and the test data. The respective values are determined as mean over ten HDNNP results and the given errors correspond to their standard deviation. The HDNNP ensembles were applied on all training and test data. $\overline{F}_{\alpha,n(E^{\prime\prime})}$ is the RMSE of QM-related atomic force components of MM atoms, which are not represented by the ML potential, but are at maximum $1\,\text{\AA}$ away from the represented atoms.
		\end{tablenotes}
		\end{center}
		\label{tab:MLP_accuracy}
	\end{threeparttable}
\end{table*}

The training accuracy for all four systems (MCL1-19G, 19G, GRP78-NKP1339, and NKP1339) is in a similar range proving the applicability of the same model parametrization for all systems. The differences in the RMSEs for the respective final training data and test data are very small for all systems (Table \ref{tab:MLP_accuracy}) revealing a good generalization of the ML potentials. 

Ensembling decreases the RMSE values for all $E_\mathrm{ML}$, $F_{\alpha,n(Q)}$, and $F_{\alpha,n(E^\prime)}$ by a few percent.

In general, the achieved accuracy is similar to those of previous approaches for ML potentials for QM/MM data \cite{Boeselt2021, Hofstetter2022}. We reach the typical target accuracy of HDNNPs for pure QM data of about $0.1\,/\,\mathrm{kJ\,mol}^{-1}$ for energies per atom and about $10\,/\,\mathrm{kJ\,mol}^{-1}\,\text{\AA}^{-1}$ for force components \cite{Behler2021}. Therefore, the approach of subtracting the electrostatic interaction between point charges of QM and MM atoms from the data to be learned seems to be justified. More importantly, the accurate representation of eeACSFs extended for QM/MM data is demonstrated. Thus, with our approach we can also construct efficient ML potentials for systems including many different chemical elements.

To examine whether a cutoff radius of $R_\mathrm{c}=5\,\text{\AA}$ is sufficient for the representation of the QM-related atomic force components of MM atoms, we calculated the RMSE of MM atoms which are between $5\,\text{\AA}$ and $6\,\text{\AA}$ away from the QM atoms. These atoms are not represented by the ML potential and hence the QM-related force prediction will be zero. We excluded atoms further away because their interaction with the QM region is smaller, thus reducing the RMSE. However, Table \ref{tab:MLP_accuracy} shows that these RMSEs of $\overline{F}_{\alpha,n(E^{\prime\prime})}$ are less than half the size of the values for $\overline{F}_{\alpha,n(E^{\prime})}$. The latter are also one order of magnitude smaller than the RMSEs for the forces of QM atoms $\overline{F}_{\alpha,n(Q)}$. Therefore, the applied cutoff radius is sufficient for the representation of QM-related forces of the MM atoms. This small cutoff radius is also a plus of the ansatz to subtract the electrostatic interaction between point charges of QM and MM atoms from the data to be learned.

\begin{figure*}[htb!]
\centering
\begin{subfigure}[h]{0.3295\textwidth}
\centering
\includegraphics[width=\textwidth]{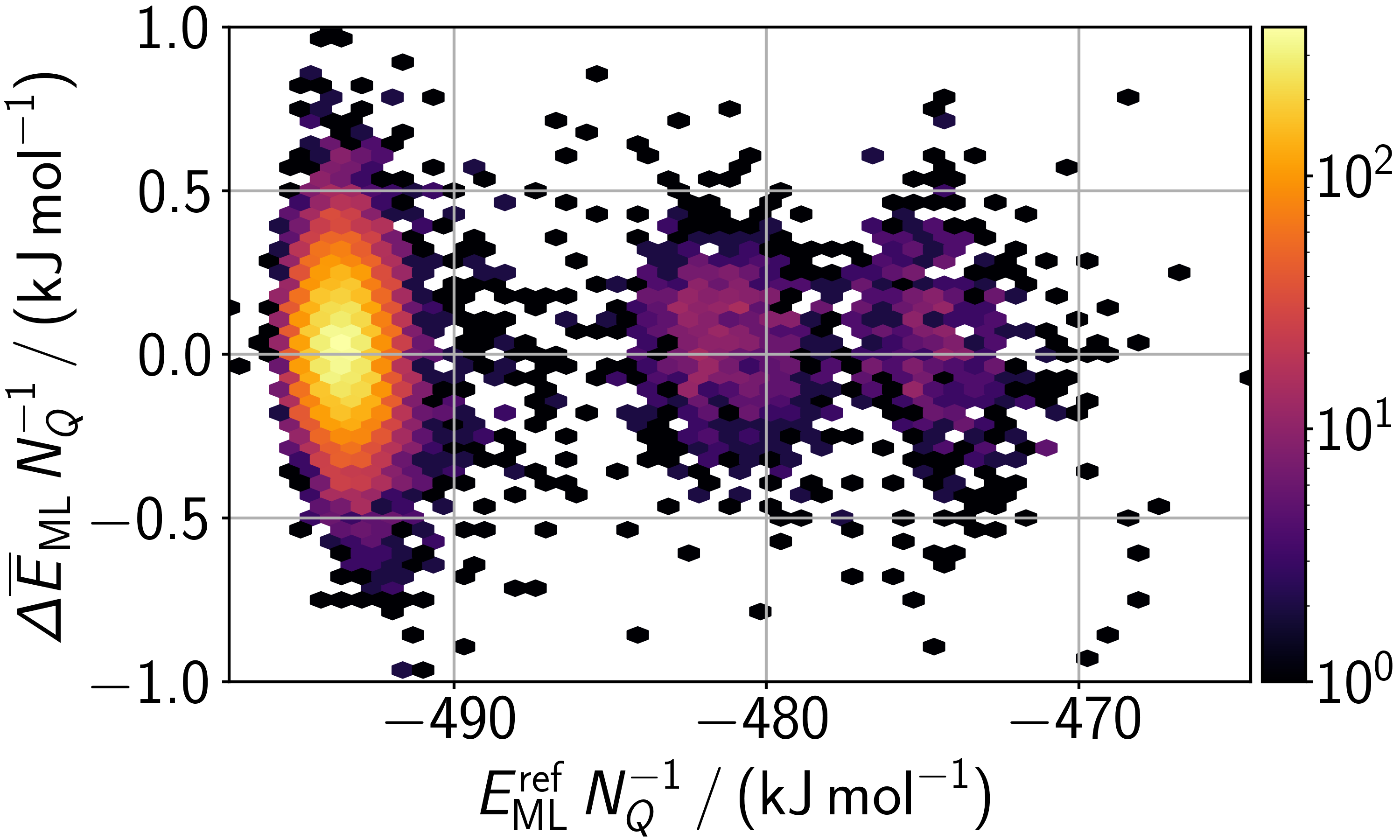}
\vspace*{-0.5cm}
\caption{}
\vspace*{0.05cm}
\end{subfigure}
\begin{subfigure}[h]{0.3295\textwidth}
\centering
\includegraphics[width=\textwidth]{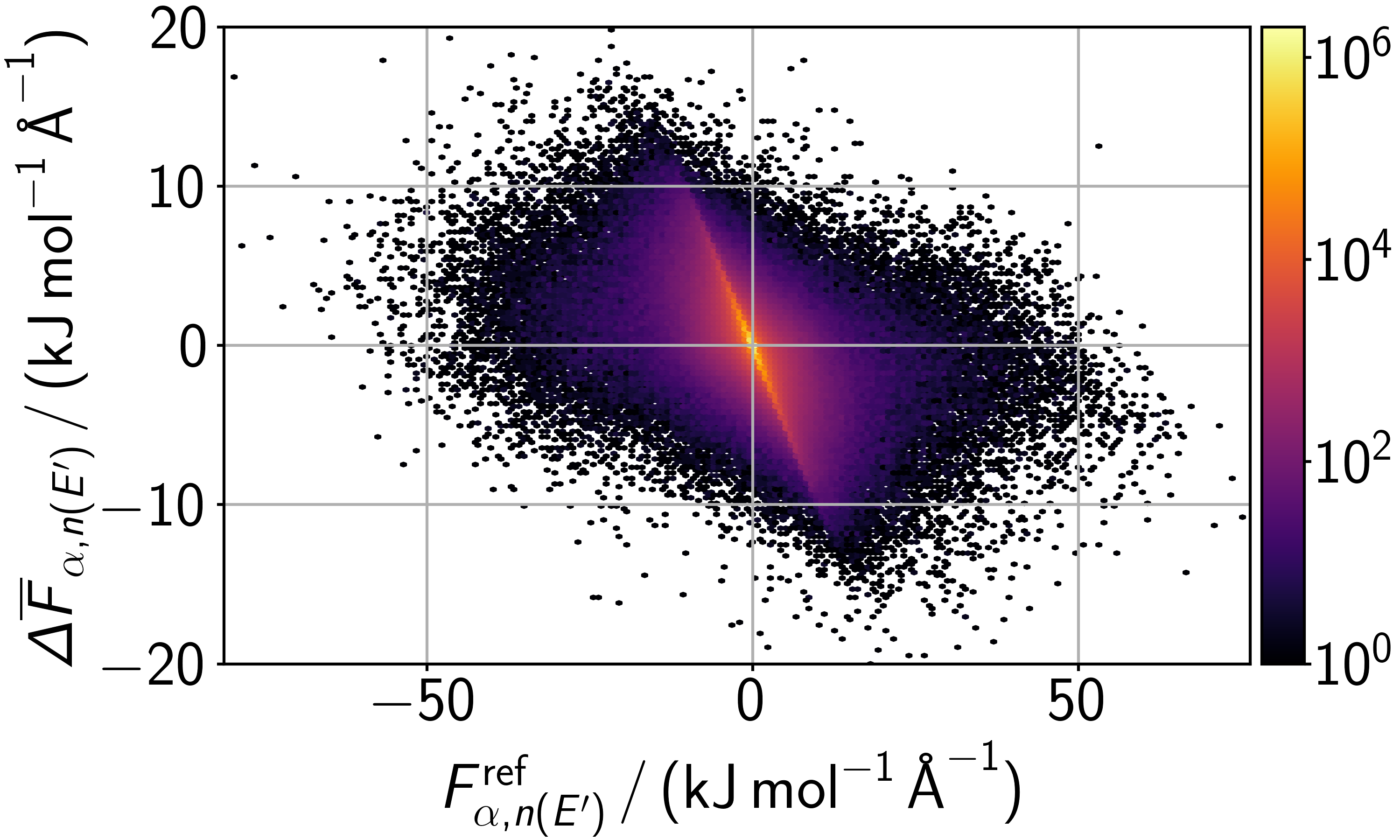}
\vspace*{-0.5cm}
\caption{}
\vspace*{0.05cm}
\end{subfigure}
\begin{subfigure}[h]{0.3295\textwidth}
\centering
\includegraphics[width=\textwidth]{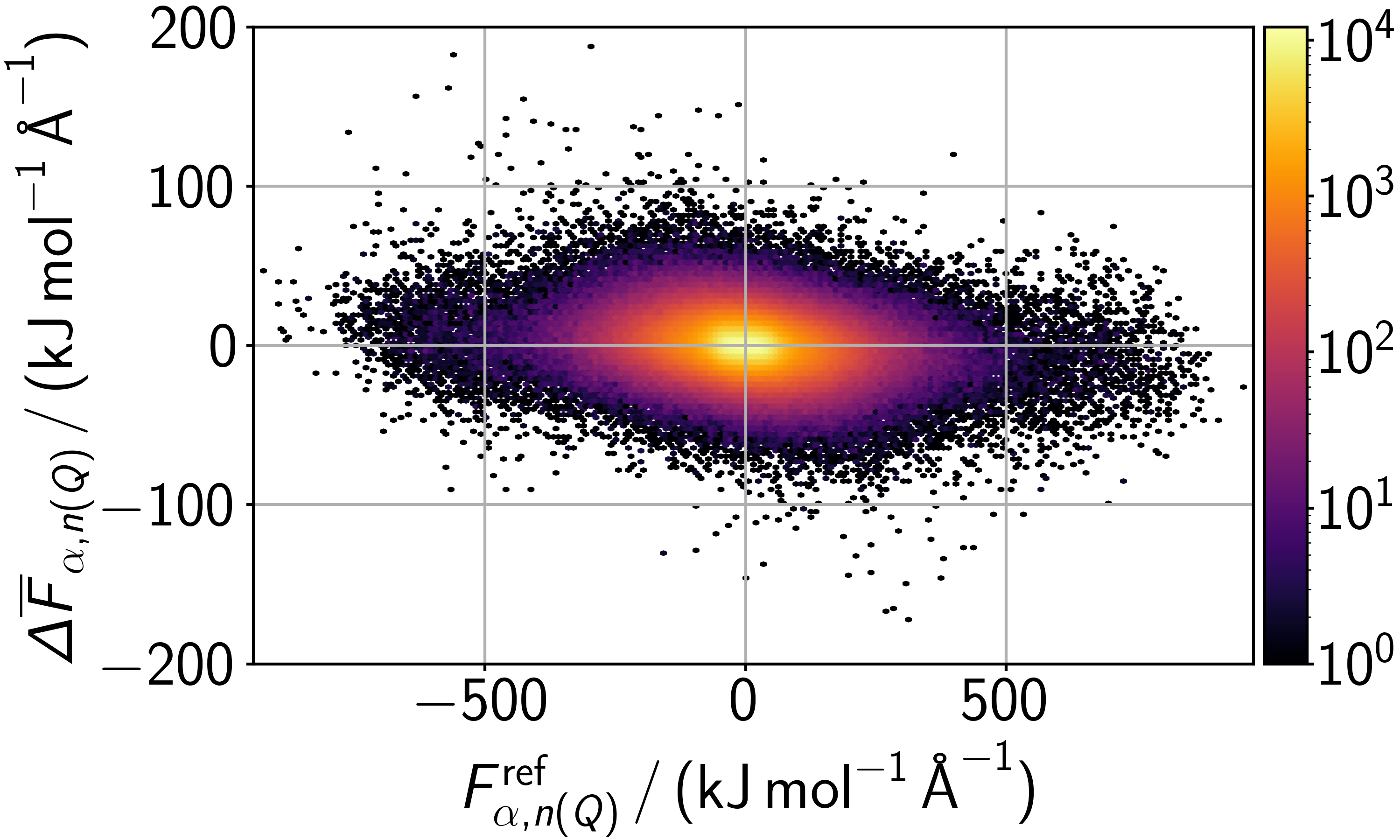}
\vspace*{-0.5cm}
\caption{}
\vspace*{0.05cm}
\end{subfigure}
\begin{subfigure}[h]{0.3295\textwidth}
\centering
\includegraphics[width=\textwidth]{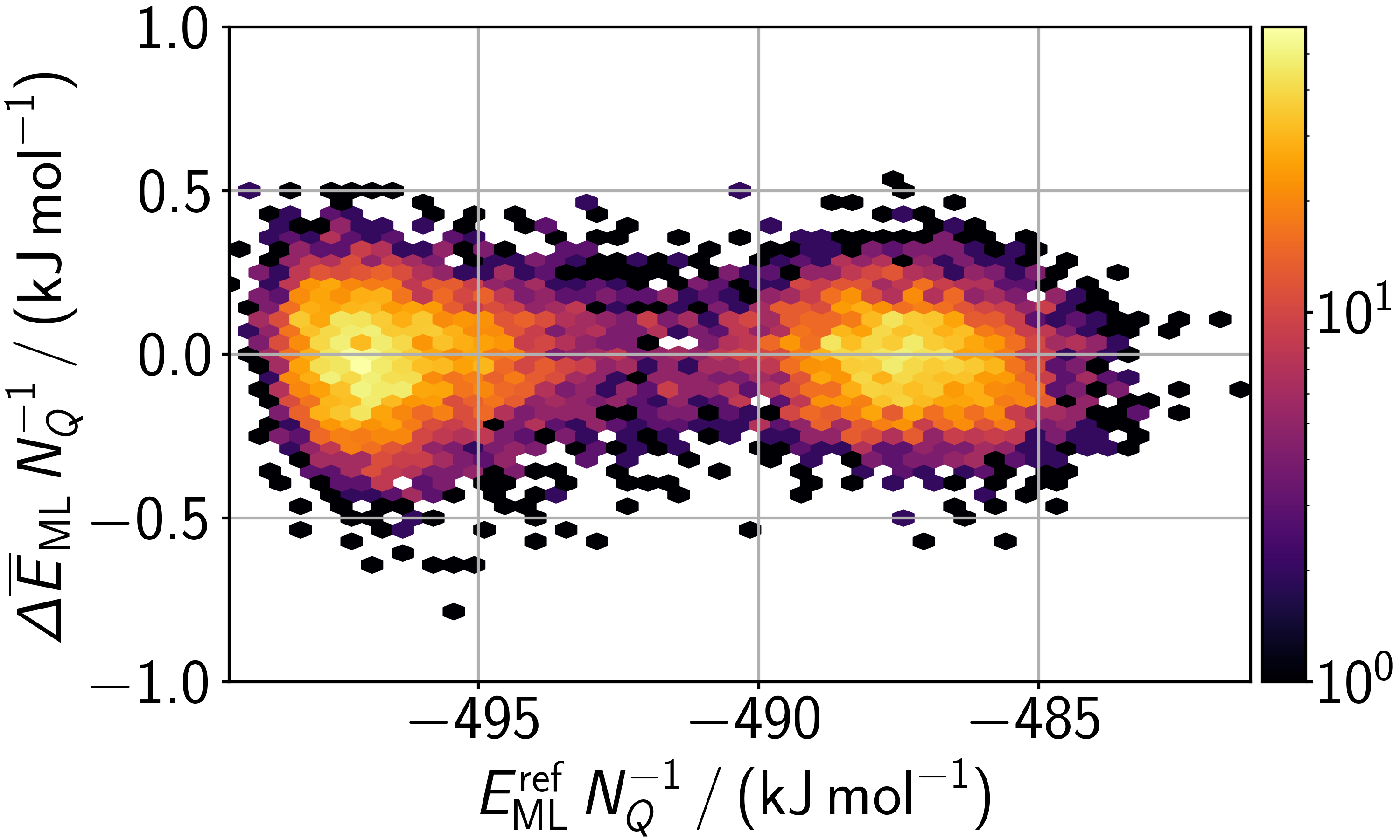}
\vspace*{-0.5cm}
\caption{}
\vspace*{0.05cm}
\end{subfigure}
\begin{subfigure}[h]{0.3295\textwidth}
\centering
\includegraphics[width=\textwidth]{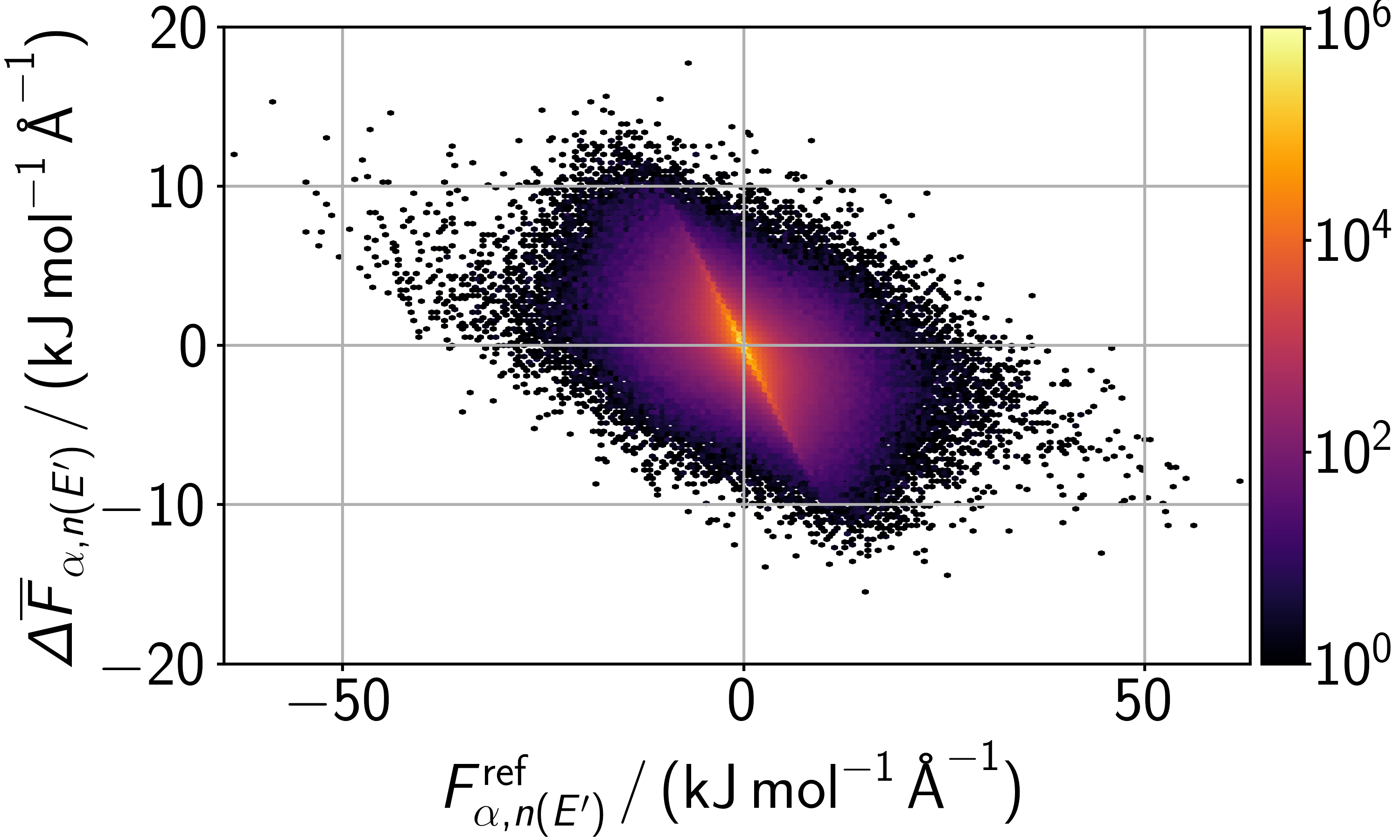}
\vspace*{-0.5cm}
\caption{}
\vspace*{0.05cm}
\end{subfigure}
\begin{subfigure}[h]{0.3295\textwidth}
\centering
\includegraphics[width=\textwidth]{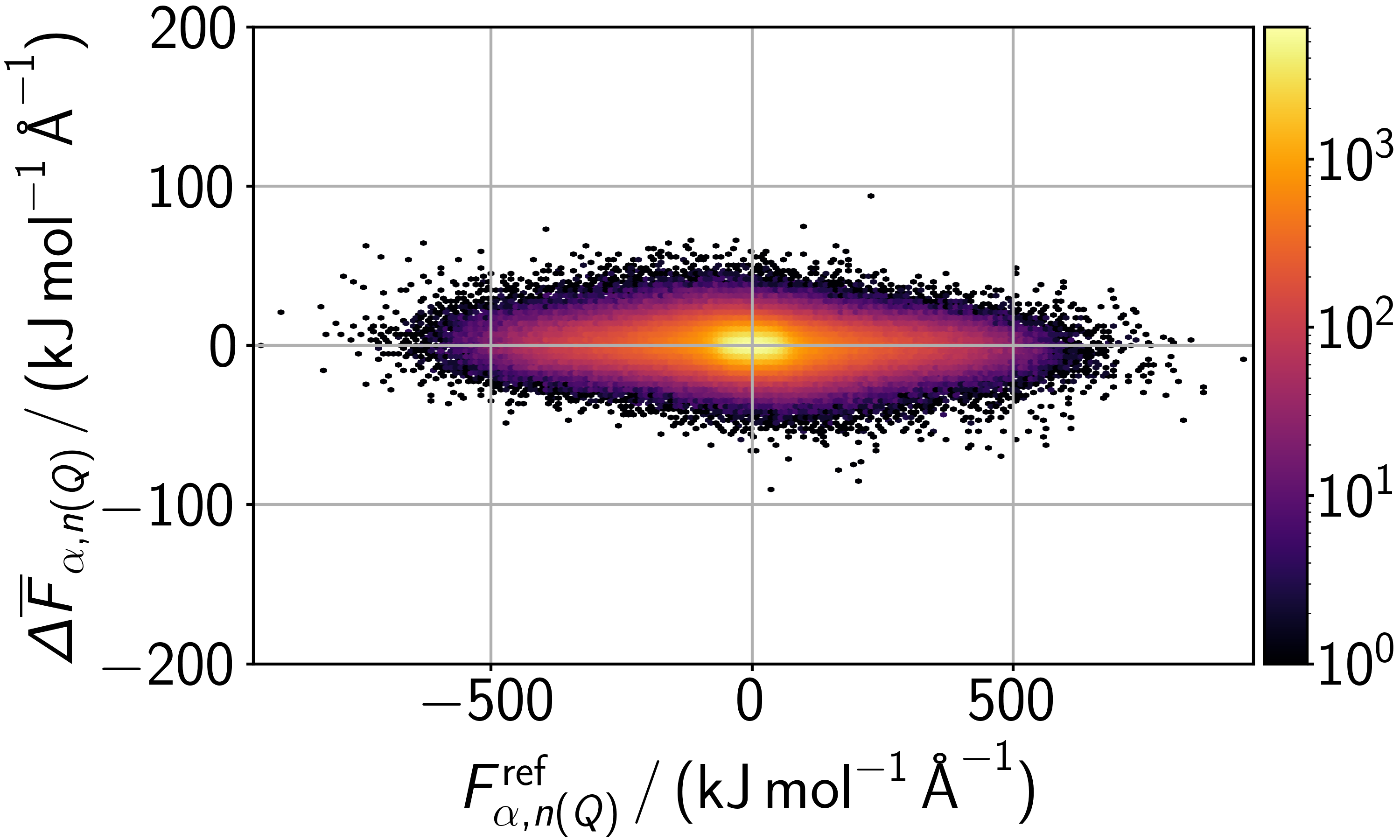}
\vspace*{-0.5cm}
\caption{}
\end{subfigure}
\caption{
Deviations between the ensemble prediction and the QM-related reference data of \textbf{(a, b, c)} both MCL1-19G and 19G and \textbf{(d, e, f)} both GRP78-NKP1339 and NKP1339. The deviations are shown for \textbf{(a, d)} energies $\Delta\overline{E}_\mathrm{ML}$ and atomic force components of \textbf{(b, e)} QM atoms $\Delta\overline{F}_{\alpha,n(Q)}$ and \textbf{(c, f)} MM atoms represented by the ML potential $\Delta\overline{F}_{\alpha,n(E^\prime)}$ as a function of the respective reference data $E_\mathrm{ML}^\mathrm{ref}$, $F_{\alpha,n(Q)}^\mathrm{ref}$, and $F_{\alpha,n(E^\prime)}^\mathrm{ref}$. The color in this hexagonal binning plot visualizes the number of data points in a hexagon. Outside the shown error ranges are \textbf{(a)} 5 and \textbf{(c)} 6 data points.
}\label{fig:MLP_error_distribution}
\end{figure*}

Figure \ref{fig:MLP_error_distribution} (a) shows the QM-related energy prediction errors of HDNNP ensembles as a function of the reference energies for the combined MCL1-19G and 19G data. Most of the reference energies are concentrated in an interval of about $4\,\mathrm{kJ\,mol}^{-1}$ per atom, which appears to be reasonable due to fluctuations in the kinetic energy at $300\,\mathrm{K}$ as well as the different chemical environments of the QM ligand atoms. The prediction errors of these data points are symmetrically centered around zero as expected. In addition, there are some high energy structures, which originate from active learning (see Figure \ref{fig:QM-region-energies}) preventing the final ML potential to drive the trajectory in this conformation space. However, for GRP78-NKP1339 and NKP1339 the reference energies are concentrated in two intervals. While the interval at higher energies includes mainly structures from the initial MM sampling, the interval at lower energies originates from active learning. Hence, the sampled conformation space by MM and ML/MM differ. This observation explains why more active learning iterations are required for GRP78-NKP1339 and NKP1339 than for MCL1-19G and 19G. Moreover, from this observation we expect that the binding free energy difference between the MM and ML/MM representation will be larger for GRP78-NKP1339 than for MCL1-19G. This is in line with the fact that MCL1-19G is a well-studied benchmark system for MM, while GRP78-NKP1339 contains a transition metal atom which is often problematic for MM.

The QM-related atomic force component error distributions for both systems look similar (Figures \ref{fig:MLP_error_distribution} (b), (c), (e), and (f)). While the force errors of the QM atoms are well centered around zero, the force errors of some part of the MM atoms show a trend for small force values. As a result of this trend, the affected forces are predicted as zero, while their reference values are slightly above or below zero. This behavior is primarily caused by MM atoms located at the edge of the cutoff sphere of certain QM atom(s), which are only described very little by the eeACSF vector. Therefore, their differentiation is difficult and the best compromise is to predict the mean value, i.e., zero. However, these force errors are below the RMSE of the force prediction error and contribute minimally to the overall error, indicating a sufficiently large cutoff radius. For larger values of $F_{\alpha,n(Q)}^\mathrm{ref}$ the prediction error distribution again centers around zero. 

%%%%%%%%%%%%%%%%%%%%%%%%%%%%%%%%%%%%%%%%%%%%%%%%%%%%%%%%%%%%%%%%%%%%%%%%%%%%%%%%%%%%%%%%%%%%%%%%%%%%%%%%
\subsection{Free Energies of Binding}
%%%%%%%%%%%%%%%%%%%%%%%%%%%%%%%%%%%%%%%%%%%%%%%%%%%%%%%%%%%%%%%%%%%%%%%%%%%%%%%%%%%%%%%%%%%%%%%%%%%%%%%%

\subsubsection*{MCL1-19G}

Our prediction of the absolute binding free energy $\Delta G^\mathrm{MM}_\mathrm{bind}$ based on the classical AFE simulations is $-37.5 \,\si{kJ.mol^{-1}} \pm 0.4\,\si{kJ.mol^{-1}}$, matching the experimental estimate ($-37.3\,\si{kJ.mol^{-1}} \pm 0.1\,\si{kJ.mol^{-1}}$) \cite{Friberg2012} nearly perfectly. Note that such close agreement with experimental data is not always the case. Previous studies using absolute AFE calculations on sets of multiple ligands have reported an RMSE to experiment of around $6-12\,\si{kJ.mol^{-1}}$ depending on the system \cite{Alibay2022, Mobley2007, Wang2006, Aldeghi2017}.

\begin{figure}[htb!]
    \centering
    \includegraphics[width=\columnwidth]{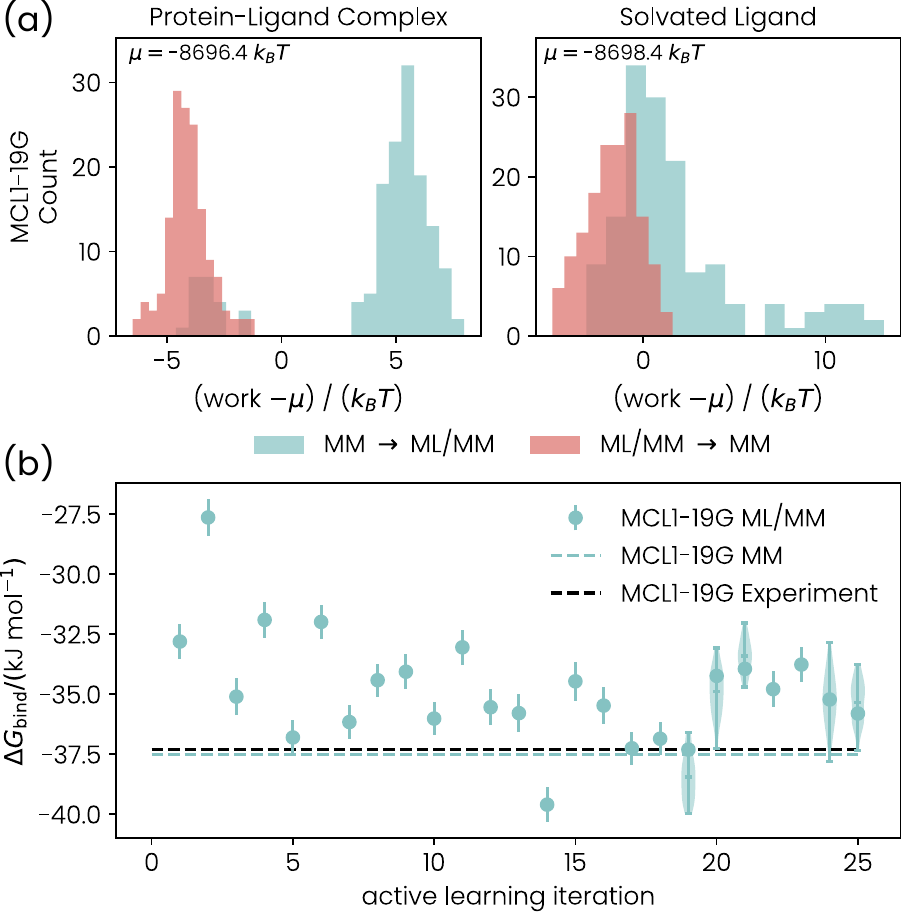}
    \caption{(a) Work distributions for the end states of MCL1-19G. The distributions were shifted by their mean $\mu$ for clarity.
             (b) Binding free energy $\Delta G_\mathrm{bind}$ as a function of the active learning iteration for MCL-19G. The error bars are the statistical error estimates from MBAR. Violin plots indicate explicit sampling of the uncertainty of the NEQ simulations. The mean and 1.5 times the interquartile range are shown in the violin plots.}
    \label{fig:work+dG.model-complex}
\end{figure}

The work distributions from the NEQ switching simulations from the MM to the ML/MM PESs for MCL1-19G are shown in Figure \ref{fig:work+dG.model-complex}(a). Overall, the work distributions for MCL1-19G are narrow, showing only a standard deviation of $3.0\,k_BT$ and $1.0\,k_BT$ for the backward and forward switches of the protein-ligand complex, respectively. We calculated the similarly low standard deviations for the work distributions of the solvated 19G ligand (forward switch: $\sigma = 3.7\,k_BT$, backward switch: $\sigma = 2.0\,k_BT$). Furthermore, the forward and backward switches show significant overlap in both cases. The low standard deviations and the overlap indicate that the relatively short switching time of $10\,\si{ps}$ was sufficient.

The work distributions of the forward switches of the solvated 19G ligand and the MCL1-19G protein-ligand complex show two distinct peaks, which disappear in the backward switches. By investigating the trajectories from the switching simulations, we found that the individual peaks correspond to different hydrogen(H)-bonding partners of the 19G ligand's carboxyl group. The peaks at low work $\mathrm{work} - \mu < 0$ are associated with H-bonds to solvent molecules, while the peak at high work is associated with an H-bond to the sulfur atom in 19G. The H-bond with sulfur is destabilized in the ML/MM compared to the MM simulation and we observe only H-bonds to solvent molecules after switching. Therefore, it requires more work to switch from MM to ML/MM if starting from a 19G conformation H-bonded to sulfur. In addition, all backward-switching simulations start from conformations with H-bonds to the solvent, which do not switch to sulfur during the simulation time, leading to a single peak for the backward switches.

To illustrate the effect of the active learning procedure on our ML/MM binding free energy estimate, we plotted $\Delta G^\mathrm{ML/MM}_\mathrm{bind}$ as a function of the active learning iteration in Figure~\ref{fig:work+dG.model-complex}(b) for MCL1-19G. Furthermore, we also show the experimental estimate and the estimate from the initial MM AFE calculations as dashed lines. We did not obtain a free energy estimate for the 0th iteration ML potential, i.e., the ML potential purely trained on the MM structures. The simulation stopped because too many structures had already been collected for active learning. Our binding free energy estimate shows a relatively wide spread of values for iterations one to six. For the higher iterations, the binding free energy estimate is scattered close to $35\,\si{kJ.mol^{-1}}$. Since the variance between iterations was significantly larger than the error estimate provided by MBAR, we sampled the binding free energy by running five additional NEQ switching simulations for each end state using the ML potential of active learning iterations 20 and 19. This sampling procedure provided 36 estimates of the binding free energy for each active learning iteration. The distributions of these estimates are illustrated as violin plots. The distributions are significantly broader than the MBAR error estimates. Furthermore, they show only little overlap, indicating that the active learning was not converged. Afterward, we increased the number of epochs for the MLP optimization from 1000 to 5000 and continued active learning up to iteration 25. Because of the high overlap between the $\Delta G^\mathrm{ML/MM}_\mathrm{bind}$ distributions for iterations 24 and 25, we considered the active learning procedure converged.

Because of the large spread of the free energy estimates, we calculated our final value of $\Delta G^\mathrm{ML/MM}_\mathrm{bind}$ as the mean of the distribution with twice its standard deviation as an error estimate $\langle\Delta G^\mathrm{ML/MM}_\mathrm{bind}\rangle = -35.3\,\si{kJ.mol^{-1}}~\pm 1.8\,\si{kJ.mol^{-1}}$. This final result is close to the experimental estimate of $-37.3\,\si{kJ.mol^{-1}} \pm 0.1\,\si{kJ.mol^{-1}}$ if we include the error ranges. Furthermore, we want to note that the protonation state of the ligand is unclear. The carboxylic acid group may be deprotonated in solution, which we did not simulate.
% Moritz: I could add here that we did this for "ease of calculation". However, we also have to compensate the molecular charge for the Ru system, which was, apparently, easy enough in that case.

Furthermore, the accuracy of our prediction workflow can be systematically improved further by multiple strategies:
(i) Currently we restricted the QM region to the ligand. However, our approach can easily extend the QM region to the protein, leading to a more accurate description of the protein--ligand interaction.
(ii) The QM region can be described with a more accurate electronic structure method, such as a hybrid exchange--correlation functional or by exploiting multilevel QM/QM embedding, as demonstrated in References \cite{Q4Bio-Paper3a} and \cite{Q4Bio-Paper3b}.

To characterize the difference between the structural ensembles produced by the MM and ML/MM potentials for the MCL1-19G system, we compared the ligand dihedral angle distributions produced by simulations with the two potentials, both in the bound and unbound state (Figure~\ref{fig:dihedral_distributions} (a)). As we did not have a converged equilibrium simulation with the ML/MM potential, we instead used an ensemble consisting of the final structure of each ML/MM equilibration simulation from the middle of the NEQ switching cycle as an approximation of the equilibrium ensemble. 
We found that the MM and ML/MM potentials produce similar dihedral angle distributions for the 19G ligand in the bound state (complex), while there were slight shifts in the populations in the unbound state (solvent). This is consistent with the narrower work distributions from NEQ switching simulations observed for the protein-ligand complex and is likely related to the higher level of ligand structural heterogeneity in the unbound state, which is also reflected in the dihedral angle distributions. Differences in the torsion potentials between the MM and ML potentials could explain why the models produce different values of $\Delta G_\mathrm{bind}$.

For comparison, we also performed end-state corrections using NEQ switching simulations from MM to an ANI-2x/MM potential, where the 19G ligand is described by ANI-2x. ANI-2x is a transferable MLP trained to reproduce QM calculations \cite{Devereux2020}. Six replicate NEQ switching calculations to the ANI-2x/MM potential resulted in $\langle\Delta G^\mathrm{ANI-2x/MM}_\mathrm{bind}\rangle = -33.14\,\si{kJ.mol^{-1}}~\pm 0.7\,\si{kJ.mol^{-1}}$. Consistent with the results using our MLP, the correction to $\Delta G_\mathrm{bind}$ is positive, although ANI-2x results in a slightly larger correction. To investigate the effect of the ANI-2x correction on the structural ensemble of the ligand, we also calculated ligand dihedral angle distributions from these simulations (Figure~S2). Interestingly, we found that the distributions were generally less perturbed by ANI-2x than our own MLP, suggesting that there is not a direct relationship between changes in ligand torsion potentials and the magnitude of the correction to $\Delta G_\mathrm{bind}$ for this system.

\subsubsection*{GRP78-NKP1339}

The results above serve as a test case where we expect a well-parameterized classical force field may perform well. We now proceed to study a more complex case involving a ligand with a transition metal. For the GRP78-NKP1339 protein--ligand complex, the classical MM AFE simulation predicts a binding free energy of $\Delta G^\mathrm{MM}_\mathrm{bind} = -19.1\,\si{kJ.mol^{-1}} \pm 1.5\,\si{kJ.mol^{-1}}$. Note that no experimental estimate is available for this protein--drug complex.

\begin{figure}[htb!]
    \centering
    \includegraphics[width=\columnwidth]{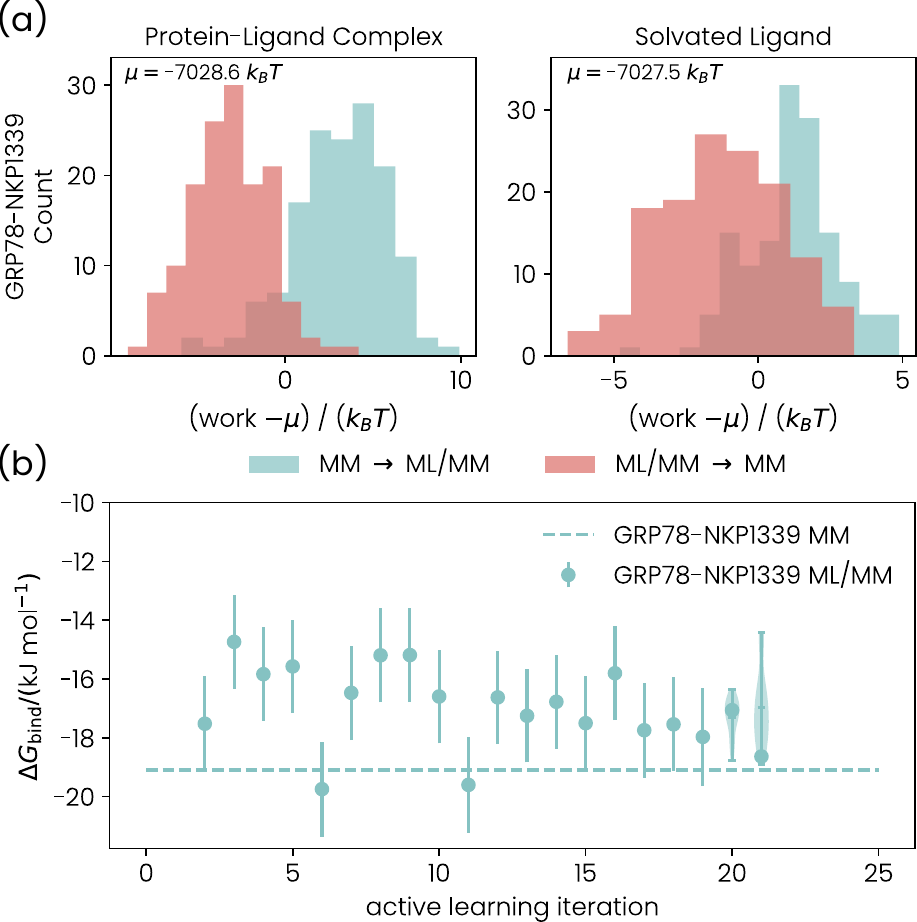}
    \caption{(a) Work distributions for the end states of GRP78-NKP1339. The distributions were shifted by their mean $\mu$ for clarity.
             (b) Binding free energy $\Delta G_\mathrm{bind}$ as a function of the active learning iteration for MCL-19G. The error bars are the statistical error estimates from MBAR. Violin plots indicate explicit sampling of the uncertainty of the NEQ simulations. The mean and 1.5 times the interquartile range are shown in the violin plots.}
    \label{fig:work+dG.ru-complex}
\end{figure}

% 1. Describe work distributions
The work distributions calculated with the final ML potential for GRP78-NKP1339 are shown in Figure~\ref{fig:work+dG.ru-complex}(a). There is significant overlap between the forward and backward work distributions for both end states (protein-ligand complex and solvated ligand). Furthermore, the distributions have only a single peak, suggesting that there is no qualitative difference between ligand conformers, which was encountered for the MCL1-19G.

In Figure~\ref{fig:work+dG.ru-complex}(b), we show the convergence of the binding free energy as a function of the active learning iteration. We did not obtain free energy estimates with the initial (purely based on MM structures) ML potential and the first ML potential from active learning. These runs did not provide binding free energy estimates because the calculations were stopped after too many structures were assigned to active learning, as discussed in Section S1. The active learning for the solvated ligand naturally converged after the 14th iteration since no additional structures were considered for training. For the protein-ligand complex, we continued the active learning until the 20th iteration, after which the number of new structures dropped below 10. We then retrained the MLP with 5000 instead of 1000 epochs and performed the final NEQ switching simulations.

To estimate the uncertainty caused by the NEQ procedure of our binding free energy estimate, we sampled the end-state corrections with five additional NEQ simulations for each end state. The resulting $\Delta G^\mathrm{ML/MM}_\mathrm{bind}$ distribution is shown as violin plots for the active learning iterations 20 and 21 in Figure~\ref{fig:work+dG.ru-complex}(b). The distribution is narrower compared to the distribution obtained for MCL1-19G. We calculated our final $\Delta G^\mathrm{ML/MM}_\mathrm{bind}$ estimate as $\langle\Delta G^\mathrm{ML/MM}_\mathrm{bind}\rangle = -17.0\,\si{kJ.mol^{-1}} \pm 2.6\,\si{kJ.mol^{-1}}$ where we used twice the standard deviation of the underlying distribution as an error. This estimate for $\Delta G_\mathrm{bind}$ differs from the MM estimate ($\Delta G^\mathrm{MM}_\mathrm{bind} = -19.1\,\si{kJ.mol^{-1}}$) by $2.1\,\si{kJ.mol^{-1}}$. However, the error bars for both results still overlap. The fact that the MM and the ML/MM estimate are close is somewhat surprising because the QM energies were significantly different for the initial MM structures and the active learning structures (see Figure~\ref{fig:QM-region-energies}. The energy gap between MM and active learning structures suggests there are qualitative differences between them. However, this large error introduced in the MM description largely cancels during the calculation of the binding free energy since it affects the solvated ligand and the protein-ligand complex.

\begin{figure*}[htb!]
\centering
    \includegraphics[width=\textwidth]{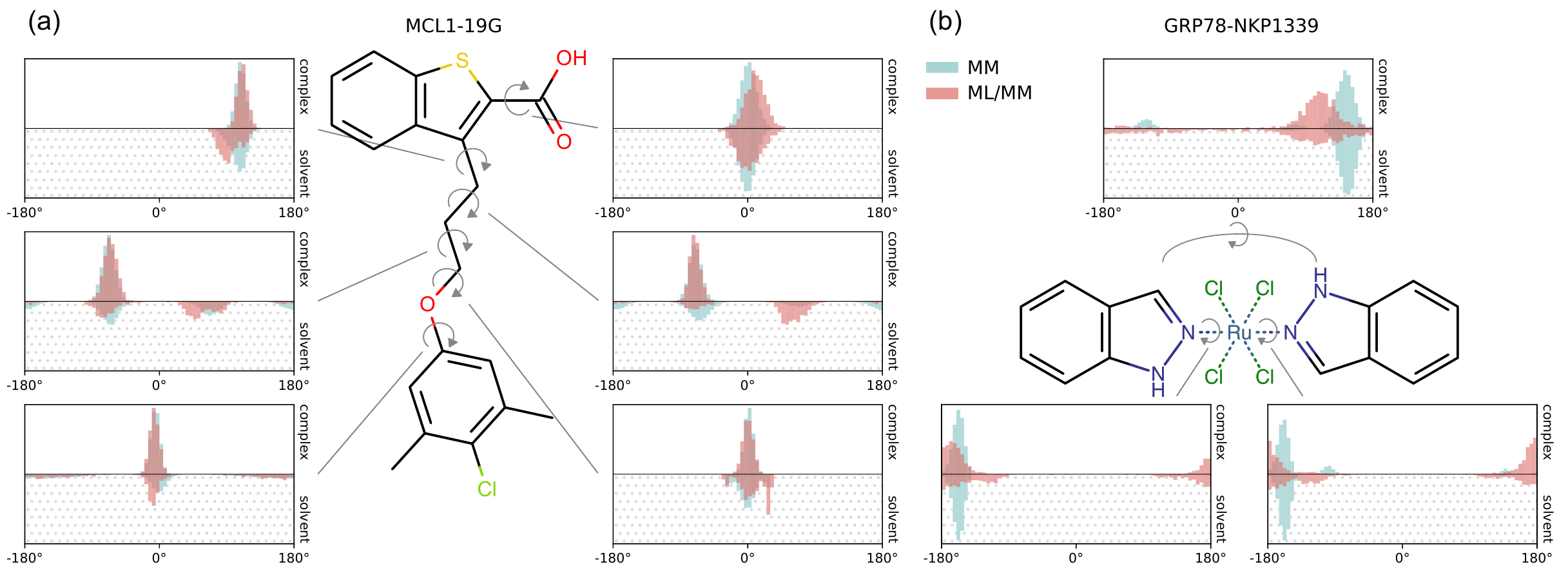}
    \caption{Dihedral angle distributions for (a) 19G in the MCL1-19G complex and in solution and (b) NKP1339 in the GRP78-NKP1339 complex and in solution. Distributions were calculated from simulations with the MM potential (blue) and ML/MM potential (red). ML/MM distributions were calculated over parallel NEQ switching simulations, using the final structure of the ML/MM equilibration step.}
    \label{fig:dihedral_distributions}
\end{figure*}

To characterize the difference between the structural ensembles produced by the MM and ML/MM potentials for the GRP78-NKP1339 system, we performed a similar analysis of ligand dihedral angle distributions as for MCL1-19G (Figure~\ref{fig:dihedral_distributions} (b)). As the indazole ring systems are planar and rigid, we focused on the dihedral angles of the indazole groups with respect to each other and the coordinated chlorides around the Ru center. We found that the ML/MM potential shifts the angle of the indazoles slightly, so they are aligned with the chlorides and closer to 90° staggered with respect to each other. In the unbound state, the ML/MM potential allows for more freedom in the rotation of the Ru-indazole bond. As for MCL-19G, these results suggest that there are differences between the MM and ML torsion potentials, which could affect the $\Delta G_\mathrm{bind}$.

%%%%%%%%%%%%%%%%%%%%%%%%%%%%%%%%%%%%%%%%%%%%%%%%%%%%%%%%%%%%%%%%%%%%%%%%%%%%%%%%%%%%%%%%%%%%%%%%%%%%%%%%
\section{Conclusions}\label{sec:conclusion}
%%%%%%%%%%%%%%%%%%%%%%%%%%%%%%%%%%%%%%%%%%%%%%%%%%%%%%%%%%%%%%%%%%%%%%%%%%%%%%%%%%%%%%%%%%%%%%%%%%%%%%%%

We presented a complete workflow to predict the absolute binding free energies of protein--ligand complexes. Our workflow employs state-of-the-art AFE simulations to predict the binding free energy using MM force fields. The MM free energies are then corrected through NEQ switching simulations, describing the work required to switch from the MM PES to a QM/MM PES. To efficiently perform the NEQ switching simulations, we used ML potentials trained explicitly for protein--ligand systems. We exploited active learning to minimize the number of QM/MM reference calculations required for our workflow and ensure that the ML potentials provided an accurate description of the QM/MM energies and forces. Furthermore, we implemented our workflow as a distributed computing framework to use the computational resources in multiple high-performance computing centers and reduce the time required to run reference calculations for the ML potentials.

During the development of the workflow, we ensured that it could be systematically improved by using more accurate electronic structure methods or multilevel QM/QM embedding, as shown in References \cite{Q4Bio-Paper3a} and \cite{Q4Bio-Paper3b}. As a second dimension to increase the workflow's accuracy, the QM region in the QM/MM embedding could be extended to the protein, increasing the accuracy of the description of the protein--ligand interaction.
Furthermore, our workflow can be applied to any ligand-protein system independent of the elementary composition. This is possible because the element embracing symmetry functions used as a descriptor in the ML potential can treat many different chemical elements efficiently, avoiding the dimensions of the descriptor growing out of proportion if complex chemical systems, such as transition metal complexes, are described. Long-range electrostatic interactions could be described by the ML potential by training it on the difference of the electrostatic interactions in QM/MM and MM models. Since the electrostatic interaction can be well described by the point-charge interaction in MM for large distances, this difference is only significant for the local environment of the QM region. Therefore, we could show that the ML potential can accurately reproduce this difference by relying on a descriptor that relies only on the local environment of each atom. We showed that our final ML approach is highly accurate, requiring only the position and elements of the system as input, and that it is highly efficient such that it can be applied in NEQ switching simulations.

We demonstrated our workflow for two protein--ligand complexes. The MCL1-19G complex served as a benchmark to establish the workflow's reliability because experimental data was available. For this system, the end-state corrections were found to be close to the already reliable results from MM. For the second test system, the complex of GRP78 and the anti-cancer drug NKP1339, we found that our initial MM parametrization provided structures with significantly higher QM energy than those obtained during active learning. The drug NKP1339 contains a Ru atom for which no established MM force field parameters were available. Therefore, the initial MM force field provided an inadequate description, which was corrected by our approach. This demonstrates that our workflow is not restricted to a specific set of elements and can be applied to a wide range of systems. Since the errors in the MM force field entered the description of the solvated ligand and the protein-ligand complex equally, the final binding free energy predicted by our ML/MM and the initial MM approach were close.

%%%%%%%%%%%%%%%%%%%%%%%%%%%%%%%%%%%%%%%%%%%%%%%%%%%%%%%%%%%%%%%%%%%%%%%%%%%%%%%%%%%%%%%%%%%%%%%%%%%%%%%%
\section*{Author Contributions}
%%%%%%%%%%%%%%%%%%%%%%%%%%%%%%%%%%%%%%%%%%%%%%%%%%%%%%%%%%%%%%%%%%%%%%%%%%%%%%%%%%%%%%%%%%%%%%%%%%%%%%%%

Moritz Bensberg: QM/MM software implementation (lead), database framework implementation (lead), force field preparation (supporting), ML descriptor implementation (lead), automatization (lead), running calculations (lead: QM/MM, ML training, NEQ switching), writing (equal).

Marco Eckhoff: Design of element-embracing atom-centered symmetry functions for QM/MM data representation, lMLP software implementation for training and prediction of QM/MM data as well as active and lifelong learning, execution of ML potential hyperparameter optimization, execution of active and lifelong learning, writing.

F.\ Emil Thomasen: AFE simulations, NEQ switching simulations for end-state corrections (supporting), structural analysis of simulations, writing.

William Bro-Jørgensen: re-implementation of the descriptor calculation in the lMLP software in C++ for speed-up, execution of QM/MM calculations and NEQ switching simulations, writing (review and editing).

Matthew S.\ Teynor: custom force field parameterization for NKP1339, lMLP-OpenMM interface implementation, NEQ switching simulations for end-state corrections, writing.

Valentina Sora: implementation of software for system preparation, execution of system preparation, writing.

Thomas Weymuth: software (supporting), force field preparation (supporting), QM/MM, MLP, NEQ calculations (supporting), writing (supporting---review and editing), supervision (supporting).

Raphael T.\ Husistein: Running ML training (supporting), ML descriptor implementation (supporting), ML potential enhancements

Frederik E.\ Knudsen: Structural analysis of simulations

Anders Krogh: Conceptualization of the project, supervision and writing (supporting --- review and editing)

Kresten Lindorff-Larsen:  Conceptualization of the project, supervision and writing (supporting --- review and editing)

Markus Reiher:  Conceptualization of the project, supervision and writing (supporting --- review and editing)

Gemma C. Solomon:  Conceptualization of the project, supervision and writing (supporting --- review and editing)

%%%%%%%%%%%%%%%%%%%%%%%%%%%%%%%%%%%%%%%%%%%%%%%%%%%%%%%%%%%%%%%%%%%%%%%%%%%%%%%%%%%%%%%%%%%%%%%%%%%%%%%%
\section*{Acknowledgments}
%%%%%%%%%%%%%%%%%%%%%%%%%%%%%%%%%%%%%%%%%%%%%%%%%%%%%%%%%%%%%%%%%%%%%%%%%%%%%%%%%%%%%%%%%%%%%%%%%%%%%%%%

This work is part of the research project ``Molecular Recognition from Quantum Computing'' and is supported by Wellcome Leap as part of the Quantum for Bio (Q4Bio) program.

Moreover, we are grateful to Novo Nordisk Foundation for financial support through the Quantum for Life center, NNF20OC0059939.
This work was also created as part of NCCR Catalysis (grant number 180544), a National Centre of Competence in Research funded by the Swiss National Science Foundation.
M.E.\ gratefully acknowledges an ETH Zurich Postdoctoral Fellowship.
Co-funded by the European Union (ERC, DynaPLIX, SyG-2022 101071843, to K.L.-L.). Views and opinions expressed are however those of the authors only and do not necessarily reflect those of the European Union or the European Research Council. Neither the European Union nor the granting authority can be held responsible for them.
G.C.S. and W.B.J. acknowledge funding from the European Research Council (ERC) under the European Union’s Horizon 2020 research and innovation programme (grant agreement No 865870). G.C.S. and M.S.T. acknowledge funding from the Novo Nordisk Foundation, Grant number NNF22SA0081175, NNF Quantum Computing Programme and NNF20OC0060019, SolidQ. AK acknowledge support from the Novo Nordisk Foundation NNF20OC0062606 and NNF20OC0063268.  

We thank Marcel Fabian for lMLP C++ code optimization and writing feedback.

\section*{Data Availability}
The machine learning potentials, the databases containing all QM/MM energies, and the software are available on ERDA\cite{erda-archive}.

\bibliography{references}

\end{document}

\title{Supporting Information: Machine Learning Enhanced Calculation of Quantum-Classical Binding Free Energies}

\renewcommand*{\thefootnote}{\dagger{footnote}}

\author{Moritz Bensberg}
\email{Authors contributed equally.}
\affiliation{ETH Zurich, Department of Chemistry and Applied Biosciences, Vladimir-Prelog-Weg 2, 8093 Zurich, Switzerland.}
\author{Marco Eckhoff}
\email{Authors contributed equally.}
\affiliation{ETH Zurich, Department of Chemistry and Applied Biosciences, Vladimir-Prelog-Weg 2, 8093 Zurich, Switzerland.}
\author{F. Emil Thomasen}
\email{Authors contributed equally.}
\affiliation{University of Copenhagen, Department of Biology, Linderstr{\o}m-Lang Centre for Protein Science, Ole Maal{\o}es Vej 5, DK-2200, Copenhagen N, Denmark.}
\author{William Bro-J{\o}rgensen}
\email{Authors contributed equally.}
\affiliation{University of Copenhagen, Department of Chemistry and Nano-Science Center, Universitetsparken 5, DK-2100, Copenhagen Ø, Denmark.}
\author{Matthew S. Teynor}
\affiliation{University of Copenhagen, Department of Chemistry and Nano-Science Center, Universitetsparken 5, DK-2100, Copenhagen Ø, Denmark.}
\affiliation{University of Copenhagen, Niels Bohr Institute, NNF Quantum Computing Programme, Blegdamsvej 21, DK-2100 Copenhagen Ø, Denmark.}
\author{Valentina Sora}
\affiliation{University of Copenhagen, Department of Computer Science, Universitetsparken 1, DK-2100, Copenhagen Ø, Denmark.}
\author{Thomas Weymuth}
\affiliation{ETH Zurich, Department of Chemistry and Applied Biosciences, Vladimir-Prelog-Weg 2, 8093 Zurich, Switzerland.}
\author{Raphael T. Husistein}
\affiliation{ETH Zurich, Department of Chemistry and Applied Biosciences, Vladimir-Prelog-Weg 2, 8093 Zurich, Switzerland.}
\author{Frederik E. Knudsen}
\affiliation{University of Copenhagen, Department of Biology, Linderstr{\o}m-Lang Centre for Protein Science, Ole Maal{\o}es Vej 5, DK-2200, Copenhagen N, Denmark.}
\renewcommand*{\thefootnote}{\fnsymbol{footnote}}

\author{Anders Krogh}
\email{akrogh@di.ku.dk}
\affiliation{University of Copenhagen, Department of Computer Science, Universitetsparken 1, DK-2100, Copenhagen Ø, Denmark and Center for Health Data Science, Department of Public Health.}
\author{Kresten Lindorff-Larsen}
\renewcommand*{\thefootnote}{\fnsymbol{footnote}}
\email{lindorff@bio.ku.dk}
\affiliation{University of Copenhagen, Department of Biology, Linderstr{\o}m-Lang Centre for Protein Science, Ole Maal{\o}es Vej 5, DK-2200, Copenhagen N, Denmark.}

\author{Markus Reiher}
\renewcommand*{\thefootnote}{\fnsymbol{footnote}}
\email{mreiher@ethz.ch}
\affiliation{ETH Zurich, Department of Chemistry and Applied Biosciences, Vladimir-Prelog-Weg 2, 8093 Zurich, Switzerland.}

\author{Gemma C. Solomon}
\renewcommand*{\thefootnote}{\fnsymbol{footnote}}
\email{gsolomon@chem.ku.dk}
\affiliation{University of Copenhagen, Department of Chemistry and Nano-Science Center, Universitetsparken 5, DK-2100, Copenhagen Ø, Denmark.}
\affiliation{University of Copenhagen, Niels Bohr Institute, NNF Quantum Computing Programme, Blegdamsvej 21, DK-2100 Copenhagen Ø, Denmark.}

\date{March 5, 2025}

\maketitle

%%%%%%%%%%%%%%%%%%%%%%%%%%%%%%%%%%%%%%%%%%%%%%%%%%%%%%%%%%%%%%%%%%%%%%%%%%%%%%%%%%%%%%%%%%%%%%%%%%%%%%%%
\section{Supporting Computational Details}
%%%%%%%%%%%%%%%%%%%%%%%%%%%%%%%%%%%%%%%%%%%%%%%%%%%%%%%%%%%%%%%%%%%%%%%%%%%%%%%%%%%%%%%%%%%%%%%%%%%%%%%%

%%%%%%%%%%%%%%%%%%%%%%%%%%%%%%%%%%%%%%%%%%%%%%%%%%%%%%%%%%%%%%%%%%%%%%%%%%%%%%%%%%%%%%%%%%%%%%%%%%%%%%%%
\subsection{System Preparation}
%%%%%%%%%%%%%%%%%%%%%%%%%%%%%%%%%%%%%%%%%%%%%%%%%%%%%%%%%%%%%%%%%%%%%%%%%%%%%%%%%%%%%%%%%%%%%%%%%%%%%%%%

We prepared the MCL1-19G and GRP78-NKP1339 systems using the Python packages pdbcraft (\url{https://github.com/Center-for-Health-Data-Science/pdbcraft}) and openmmwrap (\url{https://github.com/Quantum-for-Life/openmmwrap}) to streamline the preparation steps.

The Amber99sb*ILDN force field \cite{Lindorff‐Larsen2010} was applied to parametrize the MCL1 and GRP78 proteins. We parametrized the 19G ligand employing the GAFF 2.11 force field \cite{Wang2004, Wang2005} and the OpenFF toolkit \cite{Mobley2018}. Water was represented by the TIP3P force field \cite{Jorgensen1983}.

We generated initial force field parameters for the NKP1339 ligand by applying a system-focused atomic model without hydrogen bonding, which is implemented in SCINE Swoose \cite{SWOOSE-DEVELOP, Brunken2020, Brunken2021, Brunken2024, Weymuth2024}. For the required DFT reference calculations, PBE0-D3BJ/def2-TZVP \cite{Adamo1999, Grimme2010a, Grimme2011, Weigend2005a} was applied. We modified the resulting force field parameters to account for the square planar geometry of the chlorine atoms and the interaction between the indazole groups. Moreover, we replaced the London-type dispersion terms with standard Lennard-Jones parameters that gave the same distance and energy for the potential energy minimum.

For the MCL1-19G system, we downloaded the 4HW3 \cite{Friberg2013} structure from the Protein Data Bank \cite{Berman2000}. We removed crystallographic water and added missing heavy atoms to the protein and missing hydrogen atoms to the protein and the ligand. We capped the protein at both termini.

For the GRP78-NKP1339 system, we downloaded the 3LDO \cite{Macias2011} structure from the Protein Data Bank and kept only the protein, removing the ANP compound and crystallographic water. We added missing hydrogen atoms to the protein and capped it at both termini. Finally, we merged the protein and the handcrafted structure of the NKP1339 ligand.

Both protein-ligand systems were solvated in a box of water. Each system was first neutralized with Na$^+$ or Cl$^-$ ions, and then pairs of these ions were added to reach a salt concentration of $150\,\si{mM}$. Subsequently, we minimized the energy of each system before performing two rounds of equilibration. The first equilibration was a $2\,\si{ns}$ run in the canonical ($NVT$) ensemble employing the Langevin integrator implemented in OpenMM \cite{Izaguirre2010, Eastman2024} and harmonic position restraints for the ligand. The second equilibration was carried out for $2\,\si{ns}$ in the isothermal–isobaric ($NpT$) ensemble without restraints. The Monte Carlo (MC) barostat was applied for pressure coupling \cite{Chow1995, Aqvist2004, Eastman2024}. All equilibrations were performed at $298\,\si{K}$ with time steps of $1\,\si{fs}$.

To set up an alchemical free energy (AFE) simulation of a ligand in solution, we solvated the ligand in a box with box vectors taken from the respective equilibrated protein--ligand complex. The system was again neutralized with Na$^+$ or Cl$^-$ ions and the salt concentration was adjusted to a concentration of $150\,\si{mM}$. 

Before running AFE simulations with Yank, all systems were automatically energy minimized and equilibrated using the default settings of Yank \cite{Rizzi2019}: 1000 steps of energy minimization with the FIRE minimizer \cite{Bitzek2006} followed by one iteration of the production simulation protocol described in Section 4.4.

%%%%%%%%%%%%%%%%%%%%%%%%%%%%%%%%%%%%%%%%%%%%%%%%%%%%%%%%%%%%%%%%%%%%%%%%%%%%%%%%%%%%%%%%%%%%%%%%%%%%%%%%
\subsection{Quantum Mechanics in Molecular Mechanics Embedding}
%%%%%%%%%%%%%%%%%%%%%%%%%%%%%%%%%%%%%%%%%%%%%%%%%%%%%%%%%%%%%%%%%%%%%%%%%%%%%%%%%%%%%%%%%%%%%%%%%%%%%%%%

The quantum mechanics (QM) region in the QM/MM calculations for the training data generation was described employing Kohn--Sham density functional theory (DFT), Perdew, Burke, and Ernzerhof's exchange--correlation functional known as PBE \cite{Perdew96}, and the def2-SVP basis set \cite{Ahlrich2005}. The long-range dispersion interaction was included with Grimme's D3 dispersion correction \cite{Grimme2010a} using Becke--Johnson damping \cite{Grimme2011}. The embedded DFT calculation was polarized by the MM point charges, as described in Section 3.3. We note that only the point charges in the simulation box were included in this electrostatic embedding, and the interaction with the periodic box images was assumed to be negligible. To reduce the error of this approximation, the molecules in the simulation box were translated by the periodic box vectors such that the first atom of the QM region was at the box's center.  All DFT contributions were calculated with the quantum chemistry program Turbomole \cite{Ahlrichs1989, turbomole741}.

The MM region and the non-electrostatic interaction between the QM and MM regions was described employing the same Amber-type force field as in the system preparation. The MM contributions to the energies and forces were calculated by the program SCINE SWOOSE \cite{Brunken2020, Weymuth2024, erda-archive}.

A central database facilitated the large number of QM/MM single-point calculations required for training the ML potentials \cite{Bensberg2023m, Weymuth2024}. This database contained all molecular structures from the initial MM simulations and active learning as well as as the in- and output of the QM/MM calculations.

%%%%%%%%%%%%%%%%%%%%%%%%%%%%%%%%%%%%%%%%%%%%%%%%%%%%%%%%%%%%%%%%%%%%%%%%%%%%%%%%%%%%%%%%%%%%%%%%%%%%%%%%
\subsection{Machine Learning Potentials}
%%%%%%%%%%%%%%%%%%%%%%%%%%%%%%%%%%%%%%%%%%%%%%%%%%%%%%%%%%%%%%%%%%%%%%%%%%%%%%%%%%%%%%%%%%%%%%%%%%%%%%%%

% model parameters
We applied an ensemble size of $N_\mathrm{MLP}=10$ HDNNPs, with the uncertainty scaling factor $c=2$. The feed-forward neural networks of every element (H, C, N, O, S, Cl, and Ru) in all HDNNPs employed an input layer with $n_G=174$ neurons, three hidden layers with $n_1=120$, $n_2=72$, and $n_3=48$ neurons, and an output layer with one neuron. A scaled hyperbolic tangent \cite{Eckhoff2023} was applied as the activation function. The weights were initialized according to the scheme in Reference \cite{Eckhoff2023}. The cutoff radius of the eeACSFs was set to $R_\mathrm{c}=5\,\text{\AA}$. All other eeACSF parameter values are provided in Tables S1 and S2.

The energy contributions of neutral free atoms in their lowest spin state were removed from the QM/MM reference data for the QM atoms (see Table S3 in the Supporting Information). Structures including atomic forces higher than $10\,\mathrm{eV}\,\text{\AA}^{-1}$ were excluded before training. For each HDNNP, the reference data was split randomly into $90\%$ training structures and $10\%$ test structures. The CoRe optimizer \cite{Eckhoff2023, Eckhoff2024} (version 1.1.0) \cite{Eckhoff2024a} was applied to minimize the loss function. This loss function was equal to that in Reference \cite{Eckhoff2024} ($q=10.9$). The CoRe optimizer hyperparameters were taken from Reference \cite{Eckhoff2024}, with $\eta_-=0.55$ and $p_\mathrm{frozen}=0.025$. The weight decay and fraction of frozen weights were set as in Reference \cite{Eckhoff2023} for the weights $\alpha$ and $\beta$ and those associated with the output neuron. Lifelong adaptive data selection \cite{Eckhoff2023} was applied during training. Two updates of the weights were carried out during each training step. The first was based on $1\%$ of the current training energy predictions, and the second one relied on $1\%$ of the training energy and $1\%$ of the training force predictions of the same structures after the first update. In each training step, the selection of structures employed for the update was renewed. To obtain an HDNNP ensemble, 20 individual HDNNPs were trained and the ten best HDNNPs were selected for the ensemble. This selection was based on the sum of the test mean squared errors of energies and atomic force components, whereby the energy error was multiplied by $2500\,\text{\AA}^{-2}$. 

Active learning considered structures showing uncertainties higher than $3\cdot\mathrm{RMSE}(E_\mathrm{ML}^\mathrm{test})$, $6\cdot\mathrm{RMSE}(F_{\alpha,n(Q)}^\mathrm{test})$, and/or $18\cdot\mathrm{RMSE}(F_{\alpha,n(E^\prime)}^\mathrm{test})$ for QM-related energies and QM-related atomic force components ($\alpha=x,y,z$) of QM atoms $n(Q)$ and MM atoms within the cutoff sphere of the ML representation $n(E^\prime)$, respectively. Structures were disregarded if they included unphysical small interatomic distances. We required a minimum of 20 simulation steps between two structures chosen for active learning. For each simulation, we then selected up to eight structures. The selection was based on maximizing the simulation time separation. 150 AFE simulations were executed in parallel in each active learning iteration. During the first 20 iterations of active learning, only 1000 epochs were used during MLP training to reduce the training time. Afterward, the number of epochs was increased to 5000. The active learning procedure was considered converged when no additional structures were added to the database, or the binding free energy $\Delta G_\mathrm{bind}^\mathrm{ML/MM}$ did not change within its uncertainties. The uncertainties from the NEQ switching simulations were obtained from sampling the binding free energies from six independent NEQ simulations for each endastate, providing 36 values for $\Delta G_\mathrm{bind}^\mathrm{ML/MM}$.

% software
Training and prediction of HDNNP ensembles for QM/MM data, as well as active and lifelong learning, was implemented in the lMLP software (version 2.0.0) \cite{Eckhoff2024b}. The symmetry functions in Equations (6) %~(\ref{eq:radial_functions}) 
and (7) %(\ref{eq:angular_functions})
were implemented in an efficient C++ package 'SymmetryFunctions' providing Python and PyTorch bindings for ease of use.
All software will be made available on GitHub.

%%%%%%%%%%%%%%%%%%%%%%%%%%%%%%%%%%%%%%%%%%%%%%%%%%%%%%%%%%%%%%%%%%%%%%%%%%%%%%%%%%%%%%%%%%%%%%%%%%%%%%%%
\subsection{Alchemical Free Energy Simulations}
%%%%%%%%%%%%%%%%%%%%%%%%%%%%%%%%%%%%%%%%%%%%%%%%%%%%%%%%%%%%%%%%%%%%%%%%%%%%%%%%%%%%%%%%%%%%%%%%%%%%%%%%

Absolute binding AFE simulations were performed using Yank 0.25.2 \cite{Rizzi2019} and OpenMM 8.1.1 \cite{Eastman2024} according to the thermodynamic cycle shown in Figure 2. 
To prevent the ligand from moving through the simulation box during the protein--ligand complex dissociation simulation, we applied a harmonic restraint between the centers of mass of the protein and ligand with an equilibrium distance of zero. The spring constant was chosen such that the potential reaches $k_\mathrm{B}T$ at one radius of gyration of the protein, where $k_\mathrm{B}$ is the Boltzmann constant and $T$ is the temperature. We applied the following $\lambda$-scheme for the protein--ligand complex leg: increase the harmonic protein--ligand restraint from $\lambda=0$ to 1 in steps of $0.25$, annihilate the ligand electrostatic interactions from $\lambda=1$ to $0$ in steps of $0.1$, and finally decouple Lennard-Jones interactions between the ligand and environment from $\lambda=1$ to $0.5$ in steps of $0.1$ and from $\lambda=0.5$ to $0$ in steps of $0.05$. We employed the same $\lambda$-scheme without the harmonic restraint step for the solvation of the ligand. We ran one simulation for each $\lambda$-state. We utilized the Hamiltonian replica exchange scheme with Gibbs sampling implemented in Yank \cite{Chodera2011}, in which pairs of simulations can swap their alchemical states based on MC moves. The Metropolis criterium was

\begin{align}
\label{eq:HamiltoninanreplicaexchangeMC}
P_\mathrm{accept} = \min \biggl\{ 1,\ \frac{\exp\{-[u_{i}(x_{j})+ u_{j}(x_{i})]\}}{\exp\{-[u_{i}(x_{i})+ u_{j}(x_{j})]\}}~,
\end{align}

where $x$ is the configuration of the system in alchemical state $i$ or $j$, and $u$ is the reduced potential energy given by the force field of alchemical state $i$ or $j$. 

We propagated the simulation with the default settings in Yank. For the protein--ligand complex, simulations were run for $2 \times 10^4$ iterations employing the following protocol: (i) propose Hamiltonian replica exchange MC moves to swap the alchemical states of pairs of simulations, (ii) propose a rigid ligand displacement MC move, (iii) propose a rigid ligand rotation MC move, (iv) reassign Maxwell-Boltzmann velocities at $300\,\si{K}$, (v) $1\,\si{ps}$ of Langevin dynamics with a g-BAOAB integrator \cite{Leimkuhler2016} with a $2\,\si{fs}$ time step and $1\,\si{ps^{-1}}$ collision rate. Simulations of the ligand in solution were propagated for $2 \times 10^4$ iterations of the same protocol, but without the rigid displacement and rotation MC moves. Simulations were performed at a temperature of $300\,\si{K}$ with a MC barostat at $1\,\si{bar}$. The length of bonds to hydrogen atom was constrained. We applied the default settings in Yank for anisotropic dispersion corrections for long-range interactions, whereby the endpoints of each leg in the thermodynamic cycle are reweighted based on energies calculated with an expanded cutoff of $(0.8\,r_{\mathrm{min}})/2$, where $r_{\mathrm{min}}$ is the norm of the smallest box vector of the initial system \cite{Shirts2007}. A similar scheme is employed for the particle-mesh Ewald (PME) summation in Yank, where simulations are run with only the direct space contribution of PME, and endpoints are reweighted to take into account the reciprocal space contribution of PME. 

We analyzed the AFE simulations employing Yank. The initial simulation frames were discarded using automated equilibration detection based on the potential energies \cite{Chodera2016}. To calculate the free energy of binding (Equation (3))%(\ref{eq:total-free-energy-of-binding-1}))
, $\Delta G^\mathrm{MM}_{\mathrm{complex}}$, $\Delta G^\mathrm{MM}_{\mathrm{solv}}$, and their corresponding respective uncertainties were calculated applying the MBAR \cite{Shirts2008} as implemented in pymbar \cite{MichaelShirts2024}. $\Delta G^\mathrm{MM}_{\mathrm{complex}}$ also requires a standard state correction to account for the harmonic restraints, which corresponds to releasing the ligand from the harmonic restraint into the volume of a $1\,\si{mol.L^{-1}}$ solution. The standard state correction was calculated with radially symmetric restraints in OpenMM. The correction is given by

\begin{align}
\label{eq:dG0_from_Ka}
\Delta G_\mathrm{restraint} = -k_\mathrm{B} T \ln( C_0 V_\mathrm{L})~.
\end{align}

Here $C_0=0.6022\,\si{nm^{-3}}$ is the standard state concentration, and $V_\mathrm{L}$ is the volume of the restrained ligand. The latter is obtained by integrating the harmonic potential

\begin{align}
\label{eq:standardstatecorrection1}
V_L = \int_{0}^{r_\mathrm{max}} 4 \pi r^2 \exp\left[-\frac{1}{2}k(r-r_0)^2\right] \mathrm{d}r~.
\end{align}

The equilibrium distance is denoted by $r_0$ and $r_\mathrm{max}$ is a distance cutoff set to three times the length of the longest simulation box vector.

%%%%%%%%%%%%%%%%%%%%%%%%%%%%%%%%%%%%%%%%%%%%%%%%%%%%%%%%%%%%%%%%%%%%%%%%%%%%%%%%%%%%%%%%%%%%%%%%%%%%%%%%
\subsection{End-State Correction Simulations}
%%%%%%%%%%%%%%%%%%%%%%%%%%%%%%%%%%%%%%%%%%%%%%%%%%%%%%%%%%%%%%%%%%%%%%%%%%%%%%%%%%%%%%%%%%%%%%%%%%%%%%%%

Corrections to the end-state free energies associated with going from the MM to the QM/MM representation were determined using bidirectional NEQ switching simulations with the same approach as in References \cite{Rufa2020} and \cite{Tkaczyk2024}. We randomly sampled $150$ structures (with replacement) from the MD trajectory at the selected thermodynamic state and started a 10 ps forward switching (MM to ML/MM) simulation from each structure. In the forward switching simulations, we interpolated between the MM and ML/MM potential employing

\begin{align}
E(\lambda) = (1-\lambda) E_\mathrm{MM} + \lambda E_\mathrm{ML/MM}~,
\end{align}

where $\lambda$ is a scaling parameter that is updated to $\lambda=n/N$ every 10 time steps. $N$ is the total number and $n$ is the current time step, respectively. Every 10 steps, we calculated the dimensionless work associated with updating the potential for the current structure,

\begin{align}
\mathrm{work}_{t} = \frac{1}{k_\mathrm{B}T}[E(\lambda_{t})-E(\lambda_{t-1})]~.
\end{align}

After completion of the simulation, we calculated the accumulated work of the full switch from MM to ML/MM

\begin{align}
\mathrm{work_{MM \rightarrow ML/MM}} = \sum_t \mathrm{work}_{t}~.
\end{align}

Next, we randomly resampled with replacement 150 structures from the final snapshots of the forward switching simulations, with the probability of sampling a structure $X$ given by

\begin{align}
w(X) = \exp\left[-\mathrm{work_{MM \rightarrow ML/MM}}(X)\right]~, 
\end{align}

where $\mathrm{work_{MM \rightarrow ML/MM}}(X)$ is the accumulated work for the forward switching simulation leading to structure $X$. This approach was chosen to better approximate the equilibrium ensemble of the ML/MM representation. From each resampled structure, we ran a $10\,\si{ps}$ equilibration simulation employing ML/MM, followed by a 10 ps switching simulation from the ML/MM to the MM representation. The latter applied the same protocol as the forward switching simulation but with $\lambda=1-(n/N)$. From this backward switching simulation, $\mathrm{work_{ML/MM \rightarrow MM}}$ was obtained. Subsequently, we calculated the correction to the free energy and the associated error applying the Bennett Acceptance Ratio (BAR) between the distributions of $\mathrm{work_{MM \rightarrow ML/MM}}$ and $-\mathrm{work_{ML/MM \rightarrow MM}}$ with the pymbar package \cite{Bennett1976, Jarzynski1997, Crooks2000, Shirts2003, MichaelShirts2024}. These corrections are added to the binding free energies as shown in Equation (3).%\ref{eq:total-free-energy-of-binding-1}.

% software
The NEQ switching simulations were run with OpenMM 8.1.2 \cite{Eastman2024}. We interfaced our ML potential directly to OpenMM using the Python package 'EEForce', which will be available on GitHub. 

NEQ switching simulations with ANI-2x were performed using the TorchANI implementation in OpenMM-ML 1.2 \cite{Gao2020}. ANI-2x simulations were performed with a 1 fs time step, keeping all hydrogen constraints from the MM system setup. We used the same bidirectional NEQ switching protocol for ANI-2x endstate corrections, with the exception that $\lambda$ was updated at every time step.

\subsection{Structural analysis}

We constructed ML/MM equilibrium ensembles by taking the final structure of each ML/MM equilibration simulation from the NEQ switching cycle. The resampling step and subsequent equilibration simulations should push the distribution of structures toward the equilibrium ML/MM ensemble. We used all six NEQ switching calculations, each with 150 simulations, to construct the ML/MM ensembles. We calculated dihedral angles from these ML/MM ensembles and the endstate simulations from our MM AFE calculations using the MDAnalysis Python package \cite{Agrawal2011}. For NKP1339, we calculated the N$^\mathrm{1}$-N$^\mathrm{2}$-Ru-Cl dihedral angles for both indazoles (indazole 1 to Cl$^\mathrm{3}$ and indazole 2 to Cl$^\mathrm{4}$) and the N$^\mathrm{1}$-N$^\mathrm{2}$-N$^\mathrm{2}$-N$^\mathrm{1}$ dihedral angles between the two indazoles across the Ru center. For 19G, we calculated the C$^\mathrm{AV}$-C$^\mathrm{AW}$-C$^\mathrm{AQ}$-O$^\mathrm{AD}$, C$^\mathrm{AW}$-C$^\mathrm{AV}$-C$^\mathrm{AN}$-C$^\mathrm{AL}$,
C$^\mathrm{AV}$-C$^\mathrm{AN}$-C$^\mathrm{AL}$-C$^\mathrm{AM}$, C$^\mathrm{AN}$-C$^\mathrm{AL}$-C$^\mathrm{AM}$-O$^\mathrm{AO}$, C$^\mathrm{AL}$-C$^\mathrm{AM}$-O$^\mathrm{AO}$-C$^\mathrm{AT}$, and C$^\mathrm{AM}$-O$^\mathrm{AO}$-C$^\mathrm{AT}$-C$^\mathrm{AJ}$ dihedral angles.

%%%%%%%%%%%%%%%%%%%%%%%%%%%%%%%%%%%%%%%%%%%%%%%%%%%%%%%%%%%%%%%%%%%%%%%%%%%%%%%%%%%%%%%%%%%%%%%%%%%%%%%%
\subsection{Element-Embracing Atom-Centered Symmetry Function Parameters}
%%%%%%%%%%%%%%%%%%%%%%%%%%%%%%%%%%%%%%%%%%%%%%%%%%%%%%%%%%%%%%%%%%%%%%%%%%%%%%%%%%%%%%%%%%%%%%%%%%%%%%%%

The cutoff radius $R_\mathrm{c}=5\,\text{\AA}$ and the maximum absolute atomic charge $q_\mathrm{max}=1\,e$ were applied for all element-embracing atom-centered symmetry functions (eeACSFs).

\begin{table}[H]
\caption{Parameters of the radial eeACSFs. All combinations of the parameters in each row are applied.}
\begin{center}
\begin{tabular}{lllll}
\hline\vspace{-0.325cm}\\
$I$ && $h$ && $\eta$\vspace{0.075cm}\\
\hline\vspace{-0.325cm}\\
$t^Q,\ t^E$ && $1$ && $4.384960,\ 7.315969,\ 12.621850,\ 23.782024,\ 54.109124$\\
$t^Q$ && $n$ && $3.571065,\ 5.952354,\ 10.077045,\ 18.178648,\ 37.671588$\\
$t^Q$ && $m$ && $4.854966,\ 8.125365,\ 14.194768,\ 27.480122,\ 66.398697$\\
$t^Q$ && $\overline{n}$ && $5.374978,\ 9.039436,\ 16.026943,\ 32.018267,\ 83.174328$\\
$t^Q$ && $\overline{m}$ && $3.958754,\ 6.595853,\ 11.261492,\ 20.728746,\ 44.837797$\vspace{0.075cm}\\
\hline
\end{tabular}
\end{center}
\label{tab:radial_eeACSF_parameters}
\end{table}

\begin{table}[H]
\caption{Parameters of the angular eeACSFs. All combinations of the parameters in each row are applied.}
\begin{center}
\begin{tabular}{lllllllllll}
\hline\vspace{-0.325cm}\\
$I$ && $h$ && $\gamma$ && $\lambda$ && $\zeta$ && $\eta$\vspace{0.075cm}\\
\hline\vspace{-0.325cm}\\
$t_j^Q\cdot t_k^Q$ && $1$ && $1$ && $-1,\ 1$ && $1,\ 2,\ 6$ && $6.919852,\ 20.696185$\\
$t_j^Q\cdot t_k^Q$ && $n$ && $1$ && $-1,\ 1$ && $1,\ 2,\ 4$ && $4.687023,\ 12.059475$\\
$t_j^Q\cdot t_k^Q$ && $n$ && $-1$ && $-1,\ 1$ && $1,\ 2,\ 4$ && $5.150297,\ 13.663265$\\
$t_j^Q\cdot t_k^Q$ && $m$ && $1$ && $-1,\ 1$ && $1,\ 3,\ 7$ && $8.542166,\ 28.568769$\\
$t_j^Q\cdot t_k^Q$ && $m$ && $-1$ && $-1,\ 1$ && $1,\ 3,\ 7$ && $7.675911,\ 24.181530$\\
$t_j^Q\cdot t_k^Q$ && $\overline{n}$ && $1$ && $-1,\ 1$ && $2,\ 3,\ 10$ && $10.701335,\ 41.587164$\\
$t_j^Q\cdot t_k^Q$ && $\overline{n}$ && $-1$ && $-1,\ 1$ && $2,\ 3,\ 10$ && $9.541125,\ 34.198192$\\
$t_j^Q\cdot t_k^Q$ && $\overline{m}$ && $1$ && $-1,\ 1$ && $1,\ 2,\ 5$ && $5.670115,\ 15.575724$\\
$t_j^Q\cdot t_k^Q$ && $\overline{m}$ && $-1$ && $-1,\ 1$ && $1,\ 2,\ 5$ && $6.256063,\ 17.881461$\\
$t_j^Q\cdot t_k^E+t_j^E\cdot t_k^Q,\ t_j^E\cdot t_k^E$ && $1$ &&  && $-1,\ 1$ && $1,\ 2,\ 6$ && $4.687023$\\
$t_j^Q\cdot t_k^E+t_j^E\cdot t_k^Q$ && $n$ &&  && $-1,\ 1$ && $1,\ 2,\ 4$ && $4.687023$\\
$t_j^Q\cdot t_k^E+t_j^E\cdot t_k^Q$ && $m$ &&  && $-1,\ 1$ && $1,\ 3,\ 7$ && $4.687023$\\
$t_j^Q\cdot t_k^E+t_j^E\cdot t_k^Q$ && $\overline{n}$ &&  && $-1,\ 1$ && $2,\ 3,\ 10$ && $4.687023$\\
$t_j^Q\cdot t_k^E+t_j^E\cdot t_k^Q$ && $\overline{m}$ &&  && $-1,\ 1$ && $1,\ 2,\ 5$ && $4.687023$\vspace{0.075cm}\\
\hline
\end{tabular}
\end{center}
\label{tab:angular_eeACSF_parameters}
\end{table}

%%%%%%%%%%%%%%%%%%%%%%%%%%%%%%%%%%%%%%%%%%%%%%%%%%%%%%%%%%%%%%%%%%%%%%%%%%%%%%%%%%%%%%%%%%%%%%%%%%%%%%%%
\subsection{Element Energies}
%%%%%%%%%%%%%%%%%%%%%%%%%%%%%%%%%%%%%%%%%%%%%%%%%%%%%%%%%%%%%%%%%%%%%%%%%%%%%%%%%%%%%%%%%%%%%%%%%%%%%%%%

\begin{table}[H]
\caption{Energies of the neutral free atoms in their lowest spin state $E_\mathrm{elem}$  in Hartree ($E_\mathrm{h}$).}
\begin{center}
\begin{tabular}{lr}
\hline\vspace{-0.325cm}\\
Element & $E_\mathrm{elem}\,/\,E_\mathrm{h}$\vspace{0.075cm}\\
\hline\vspace{-0.325cm}\\
H & $-0.498629930$\\
C & $-37.751244825$\\
N & $-54.466595225$\\
O & $-74.914700682$\\
S & $-397.800714902$\\
Cl & $-459.797166452$\\
Ru & $-94.792756987$\vspace{0.075cm}\\
\hline
\end{tabular}
\end{center}
\label{tab:atom_energies}
\end{table}

%%%%%%%%%%%%%%%%%%%%%%%%%%%%%%%%%%%%%%%%%%%%%%%%%%%%%%%%%%%%%%%%%%%%%%%%%%%%%%%%%%%%%%%%%%%%%%%%%%%%%%%%
\section{Supporting Results}

\subsection{High Molecular Mechanics Energy Structures}

\begin{figure}[H]
    \centering
    \includegraphics[width=0.25\textwidth]{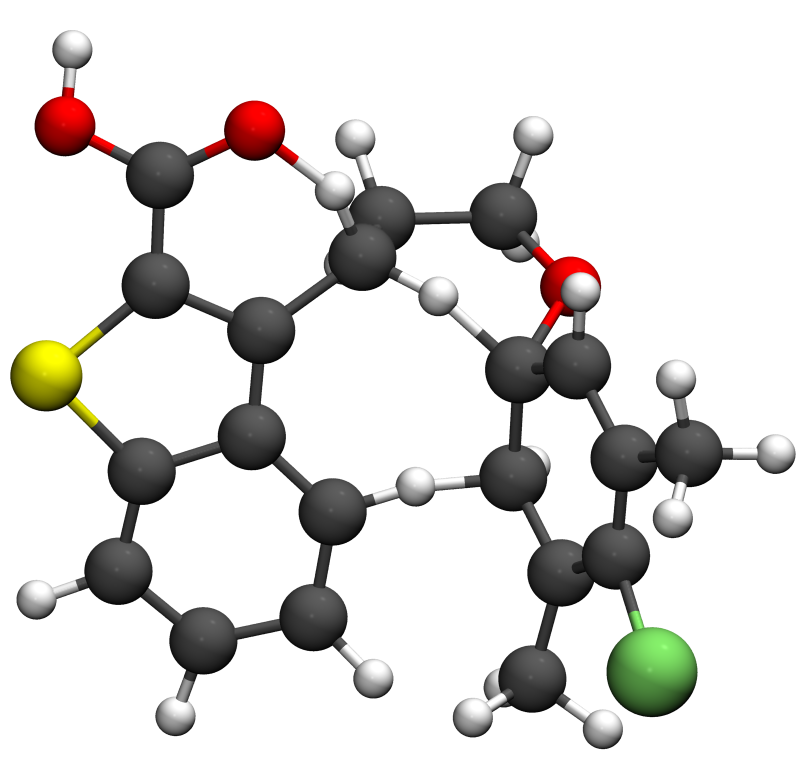}
    \caption{Structure from active learning showing a very high MM energy and unusually short C-H-C and O-H-C bonds.}
    % Energy: -1854.84427 (difference to median: -819.23 kJ/mol)
    \label{fig:high-mm-energy-structure}
\end{figure}

During active learning of the MCL1-19G complex and the solvated 19G ligand, structures were encountered that showed high molecular mechanics (MM) energies ($\sim 400~\si{kJ.mol^{-1}}$ to $900~\si{kJ.mol^{-1}}$ higher than the median). These high energies are caused by short bond distances that MM force fields cannot accurately describe. Figure \ref{fig:high-mm-energy-structure} shows an example of such a structure. This structure has an unusually short O--H bond between atoms that are not covalently bonded according to the MM topology. Such a transition state-like conformation can be handled by quantum mechanics (QM) but not by MM.

\subsection{Endstate corrections with ANI-2x}

\begin{figure}[H]
    \centering
    \includegraphics[width=\textwidth]{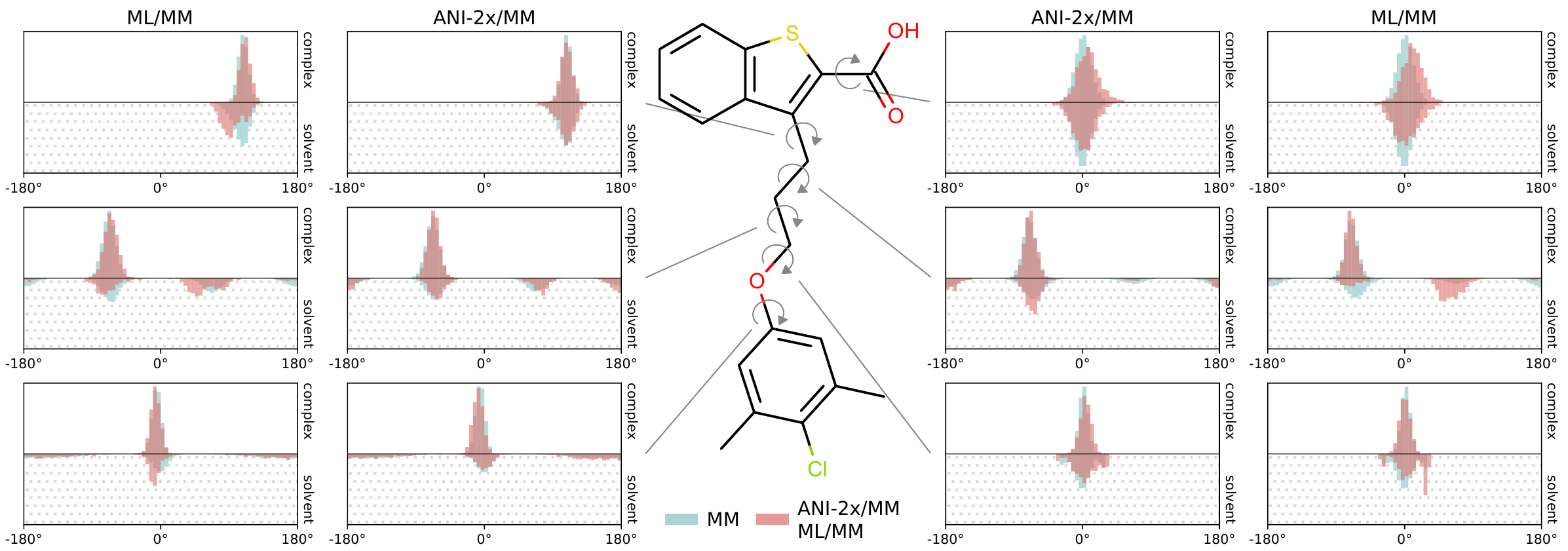}
    \caption{Comparison of dihedral angle distributions produced by our MLP and ANI-2x for 19G in the MCL1-19G complex and in solution. Distributions were calculated from simulations with the MM potential (blue) and ANI-2x/MM or our own ML/MM potential (red). The columns of plots are labeled according to the force field. ANI-2x/MM and ML/MM distributions were calculated over parallel NEQ switching simulations, using the final structure of the equilibration step.}
    \label{fig:ani2x_dihedral_distributions}
\end{figure}

\bibliography{references}